%% file: iccp24_template.tex
\pdfoutput=1
\documentclass[10pt,journal,compsoc]{IEEEtran}
\newif\ifpeerreview

\peerreviewfalse

\newcommand{\rev}[1]{\textcolor{black}{ #1}}

\newcommand*\diff{\mathop{}\!\mathrm{d}}

\usepackage[nocompress]{cite}
\usepackage{url}
\usepackage{amsmath,amssymb,graphicx}
\usepackage{lipsum} %
\usepackage[switch]{lineno}
\usepackage{lineno}
\usepackage{threeparttable}
\usepackage{etoc}
\usepackage{multirow}
\usepackage{float}
\usepackage{gensymb}
\usepackage{graphicx}
\usepackage{caption}
\usepackage{flushend}
\usepackage{array} 
\usepackage[table,RGB]{xcolor}
\usepackage{booktabs}
\usepackage{algorithm}%
\usepackage{algpseudocode}%
\usepackage{hyperref}

\usepackage{xcolor} 
\definecolor{Colorfmod}{RGB}{100, 131, 245} %
\definecolor{Colorfdet}{RGB}{251, 161, 60} %
\definecolor{Colormfo}{RGB}{211, 82, 48} %
\definecolor{CustomBlue2}{RGB}{91, 129, 200}
\definecolor{CustomGreen}{RGB}{7, 179, 88}

\newcommand{\paperID}{XXXX}

\title{High-performance real-world optical computing trained by in situ gradient-based model-free optimization}
\author{Guangyuan~Zhao,~\IEEEmembership{Student Member,~IEEE,}
        Xin~Shu,~\IEEEmembership{Member,~IEEE,}
        and~Renjie~Zhou,~\IEEEmembership{Member,~IEEE}%
        
\IEEEcompsocitemizethanks{
\IEEEcompsocthanksitem{This work is supported in part by Hong Kong General Research Fund 14209521, in part by Hong Kong Innovation and Technology Fund ITS/178/20FP \& ITS/148/20 and in part by Croucher Foundation CM/CT/CF/CIA/0688/19ay. (G. Zhao and X. Shu contributed equally to this work.) (Corresponding author: Guangyuan Zhao.)}
\IEEEcompsocthanksitem G. Zhao, X. Shu and R. Zhou are with the Department
of Biomedical Engineering, the Chinese University of Hong Kong, Hong Kong, China (email: zhaoguangyuan@link.cuhk.edu.hk, shuxin@link.cuhk.edu.hk, rjzhou@cuhk.edu.hk).
}%
}

\begin{document}

\IEEEtitleabstractindextext{%
\begin{abstract}
Optical computing systems provide high-speed and low-energy data processing but face deficiencies in computationally demanding training and simulation-to-reality gaps. We propose a gradient-based model-free optimization (G-MFO) method based on a Monte Carlo gradient estimation algorithm for computationally efficient in situ training of optical computing systems. This approach treats an optical computing system as a black box and back-propagates the loss directly to the optical computing weights' probability distributions, circumventing the need for a computationally heavy and biased system simulation. Our experiments on diffractive optical computing systems show that G-MFO outperforms hybrid training on the MNIST and FMNIST datasets. Furthermore, we demonstrate image-free and high-speed classification of cells from their marker-free phase maps. Our method's model-free and high-performance nature, combined with its low demand for computational resources, paves the way for accelerating the transition of optical computing from laboratory demonstrations to practical, real-world applications. 
\end{abstract}

}

\maketitle

\input{sections/main_onn_submission}

\ifpeerreview \else
\section*{Acknowledgments}
We thank Dr.~\href{https://zcshinee.github.io/}{Cheng Zheng} from MIT for the discussions in the early stages of the work.

\noindent \textbf{Author Contributions}. G.Z. conceived the project, derived
the formulation, and built the backbone code and system.
X.S. helped with the concept development, code writing and system setup and
collected the simulation and experiment results. R.Z. supervised the project, acquired the funding support for the project and proposed the application for label-free real-time cell analysis. G.Z. and X.S. wrote the manuscript with
comments and edits from R.Z..
\fi

\bibliographystyle{IEEEtran}
\bibliography{iccp24_template}

\newpage
\begingroup
\newpage
\onecolumn

\input{sections/osa-supplemental-document-template}
\endgroup

\end{document}

%% file: sections/main_onn_submission.tex
\IEEEraisesectionheading{\section{Introduction}}

\IEEEPARstart{O}{ptical} computing leverages the properties of light waves to facilitate high-speed data processing while reducing the energy cost~\cite{marechal1953filtre, cutrona1960optical,o1956spatial, ambs2010optical, wetzstein2020inference, lugt1964signal, zhou2019optical}. Recent advances in automatic differentiation have enabled in silico training of large-scale optical computing weights, giving rise to the realizations of diffractive neural networks~\cite{chang2018hybrid, lin2018all}, optical reservoir computing~\cite{rafayelyan2020large, verstraeten2007experimental}, and coherent nanophotonic circuits~\cite{shen2017deep}.

\textit{Problem Statement.} Training optical computing systems presents two challenges: an intensive computational process and a performance disparity between simulation and reality when implementing pre-trained weights onto real-world systems~\cite{buckley2023photonic, lin2018all, rumelhart1986learning}.
Typically, the optical computing systems are trained in silico using differentiable simulators rooted in the first principle of optics, an approach known as simulator-based training (SBT). While SBT has proven effective within the confines of the simulator, the performance in real systems is largely contingent upon the simulator's fidelity. Factors such as misalignment and aberration, often omitted in simulations, cause significant performance degradation when optical computing weights trained exclusively within the simulator are applied to real-world systems.

To bridge the reality gap between simulation and experiments, physics-aware training (PAT) and hybrid training (HBT) have been introduced ~\cite{wright2022deep, spall2022hybrid} recently. Both training strategies include conducting the forward pass in the real-world system and back-propagating the loss from the system to its weights through the simulator. These in situ approaches allow the training process to access the optical computing system during the forward pass, which leads to more accurate weight updates than in silico training ~\cite{buckley2023photonic}.

Despite these recent advances, there is a continued reliance on a physics-based simulator during the back-propagation process in current in situ training methods. Such a setting brings three drawbacks: (1) the bias between the simulator and real system prohibits the above training process from achieving optimal results; (2) the in silico simulation requires large memory and computation, limiting the aforementioned methods from in situ training in edge devices with limited computing resources~\cite{sludds2022delocalized}; and (3) the model-based training strategies need high-fidelity images of the input object, which are costly to acquire in terms of instruments, time and storage memory. Hence, developing a memory- and computation-efficient algorithm to train the optical computing system efficiently is an open problem we address in this paper.

\begin{figure}[ht]
\centering
\includegraphics[width=\linewidth]{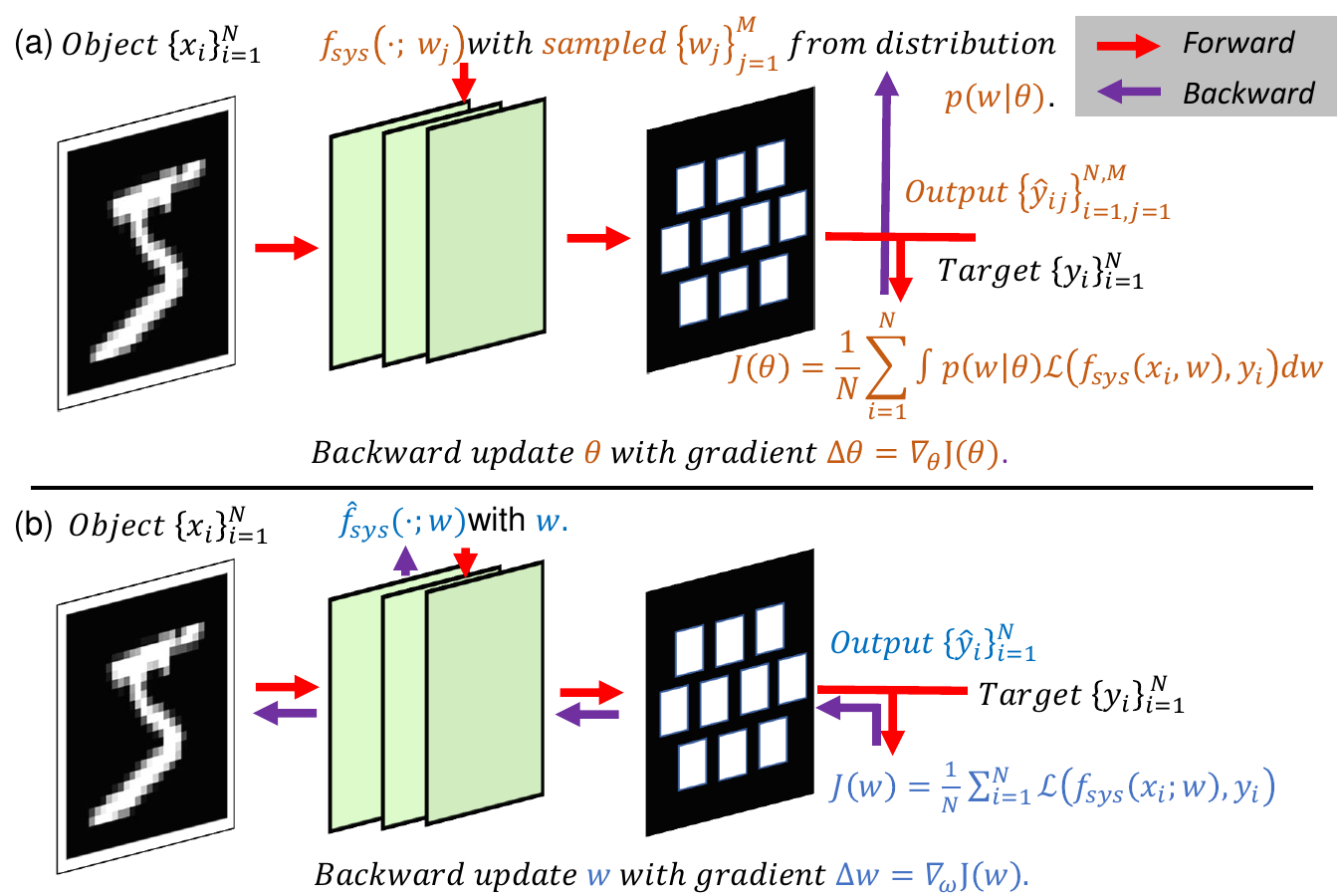}
\caption{\textbf{Gradient-based model-free optimization based training of the optical computing system.} (a) The \textcolor{Colormfo}{brown} highlights show that our gradient-based model-free training strategy back-propagates the error of training to the distribution parameter $\theta$ and bypasses the reliance on correct differentiable modeling of the optical system $f_{sys}$ and knowledge on input $\{x_i\}_{i=1}^N$.
(b) The \textcolor{CustomBlue2}{blue} highlights show that the conventional training of the optical computing system relies on a physics-based simulator $\hat{f}_{sys}$, which substitutes the inaccessible $f_{sys}$ corresponding to the real system. The training process back-propagates the loss through simulator $\hat{f}_{sys}$ to update the weight $w$. This is the basis of SBT and HBT methods.
}
\verb||\label{fig:mfo}
\end{figure}

\textit{Proposed Solution.} Here, we propose an in situ model-free optimization method that does not require back-propagating errors through the simulator. Instead, we use a score gradient estimation algorithm~\cite{wierstra2014natural, williams1992simple} to solely use the forward outputs from the real system to get gradients for updating the weights of the optical computing system. As shown in Fig.~\ref{fig:mfo}, our method treats the optical system as a black box and back-propagates task-specific negative-loss as rewards to the source weight distributions (Fig.~\ref{fig:mfo}a). This process only requires knowledge of the weights and forward outputs of the optical computing system, unlike the SBT and HBT methods that require a simulator and the images of the input objects (Fig.~\ref{fig:mfo}b).

\textit{Contributions.} We make the following contributions:

\begin{itemize}
    \item We \textit{introduce} a score-gradient estimation method to train the optical computing system in a model-free manner, providing a computation- and memory-efficient way to mitigate the reality gap to training the real optical computing system (Sec.~\ref{sec: method}). 
    \item We \textit{validate} our method on a diffractive optical computing system. Experimental results show that G-MFO outperforms hybrid training on the commonly used MNIST and FMNIST datasets~\cite{lin2018all, wright2022deep, spall2022hybrid} (Sec.~\ref{subsec: general performance}). 
    \item We \textit{show} the G-MFO training process only consumes $\sim 0.1\%$ of the GPU time and memory compared with the HBT in situ method by avoiding the computation-intensive modeling of the wave propagation process (Sec.~\ref{subsec: source efficient training by G-MFO.}).
    \item Application scenario: We \textit{demonstrate}, as a proof-of-concept, that our G-MFO-trained optical computing system's effectiveness on 
     classifying four types of white blood cells from their phase maps in a \textit{marker-free}\footnote{
    Herein, we use "marker-free" but not "label-free" to avoid confusion. This is because "label-free" has different meanings in biomedical research and high-level vision.} manner with a testing accuracy of $73.8\%$, making it a promising approach for image-free, marker-free, and high-speed cell analysis (Sec.~\ref{subsec: classify WBC.}).
\end{itemize}

\textit{Open source.} Our code will be available at repository
\href{https://github.com/shuxin626/Model-free-Computational-Optics}{Model-free-Computational-Optics} to facilitate further research and applications on computational optics and training of optical computing systems. We hope to promote the reproducibility and follow-ups for this hardware-centric research direction.

\begin{table*}[ht]
\centering
\arrayrulecolor{black}
\fontsize{8pt}{8pt}\selectfont
    \centering
    \begin{tabular}{c|p{1.6cm}|p{1.3cm}|p{1.3cm}|p{1.3cm}|p{1.3cm}|p{1.3cm}|p{1.3cm}|p{1.3cm}}
    \toprule
    & Conventional &\multicolumn{2}{c|}{MBO} & \multicolumn{5}{c}{MFO} \\\midrule
    & SBT ~\cite{lin2018all, shen2017deep} & H-MBO ~\cite{wright2022deep, spall2022hybrid, zhou2021large} & L-MBO \cite{zheng2023dual, huo2023optical} & IMOB \cite{Zhou:2020pr, nakajima2022physical, hughes2018training}  & FFT ~\cite{momeni2023backpropagation, oguz2023forward} & Trial and Error~\cite{bueno2018reinforcement, liu2022programmable} &  GA ~\cite{zhang2021efficient} & \textbf {G-MFO (ours)}  \\\midrule
    In situ & \hfil\cellcolor{red!20} No & \hfil\cellcolor{green!20} Yes & \hfil\cellcolor{green!20}Yes & 
    \hfil\cellcolor{green!20}Yes & \hfil\cellcolor{green!20}Yes & \hfil\cellcolor{green!20}Yes & \hfil\cellcolor{green!20}Yes & \hfil\cellcolor{green!20}Yes \\
 Computation overhead & \hfil\cellcolor{red!20} High & \hfil\cellcolor{red!20} High & \hfil\cellcolor{red!20} High &  \hfil\cellcolor{green!20} Low &  \hfil\cellcolor{green!20} Low &  \hfil\cellcolor{green!20} Low &  \hfil\cellcolor{green!20} Low &  \hfil\cellcolor{green!20} Low\\ 
     Input image free & \hfil\cellcolor{red!20} No & \hfil\cellcolor{red!20} No & \hfil\cellcolor{red!20}No & 
    \hfil\cellcolor{green!20}Yes & \hfil\cellcolor{red!20}No & \hfil\cellcolor{green!20}Yes & \hfil\cellcolor{green!20}Yes & \hfil\cellcolor{green!20}Yes \\
    In silico training time & \hfil\cellcolor{red!20} High & \hfil\cellcolor{red!20} High & \hfil\cellcolor{red!20} High &  \hfil\cellcolor{green!20} Low &  \hfil\cellcolor{green!20} Low &  \hfil\cellcolor{green!20} Low &  \hfil\cellcolor{green!20} Low &  \hfil\cellcolor{green!20} Low\\ 
    Dimensionality & \hfil\cellcolor{green!20} High & \hfil\cellcolor{green!20} High & \hfil\cellcolor{green!20} High &  \hfil\cellcolor{green!20} High &  \hfil\cellcolor{green!20} High &  \hfil\cellcolor{red!20} Low &  \hfil\cellcolor{red!20} Low &  \hfil\cellcolor{orange!20} Medium\\
    Intermediate measurement & \hfil\cellcolor{green!20} No & \hfil\cellcolor{green!20} No & \hfil\cellcolor{green!20}No & \hfil\cellcolor{red!20}Yes & \hfil\cellcolor{red!20} Yes & \hfil\cellcolor{green!20}No & \hfil\cellcolor{green!20}No & \hfil\cellcolor{green!20}No\\
    Gradient & \hfil\cellcolor{green!20} Yes & \hfil\cellcolor{green!20} Yes & \hfil\cellcolor{green!20} Yes & 
    \hfil\cellcolor{green!20}Yes & \hfil\cellcolor{red!20}No & \hfil\cellcolor{red!20}No & \hfil\cellcolor{red!20}No & \hfil\cellcolor{green!20}Yes\\
    \bottomrule
    \end{tabular}
    \caption{{\textbf{Comparison of strategies on training optical computing systems} along the axes of in situ training capability (in situ), in silico computation overhead (computation overhead), the requirement on the recording of the input object to the task (input image free), required in silico training time, the dimensionality of trainable parameters (dimensionality), requirements on intermediate measurement and whether or not use gradient-based update that is more efficient (gradient). 
    Simulator-based training (SBT), Hybrid model-based training (H-MBO), Learned model-based training (L-MBO), Intermediate measurement optical backpropagation (IMOB), Forward-forward training (FFT), Trial and error, Genetic algorithm (GA), Gradient-based model-free optimization (G-MFO).
    }}
    \label{tab: comparison of training methodlogies}
\vspace{-1em}
\end{table*}

\section{Related Work}

We discuss the following work related to our contributions.

\subsection{In situ training strategies to optimize the optical computing system}

Most research into optical computing system training has historically relied on in silico training. This process involves forward and backward propagation calculations on an external computer that simulates the physical system through a digital twin~\cite{lin2018all, shen2017deep}. However, this technique can lead to discrepancies between the simulation and actual reality due to inaccuracies in the physical system's representation.

In recent years, progress has been made in developing in situ training strategies that use data gathered from real-world optical computing systems to mitigate the reality gap and improve experimental performance. As a first attempt, we categorize previous and our in-situ training methods into two genres: model-based optimization (MBO) and model-free optimization (MFO) methods. We differentiate these two genres by whether they use the optical computing system emulator models during the parameter updating process. 
A tabular comparison of the previous and our methods is in Tab.~\ref{tab: comparison of training methodlogies}. Details of the illustrations and comparisons are below. 

\textbf{MBO methods.} \textit{Physics-aware training (PAT)~\cite{wright2022deep}, hybrid training (HBT)~\cite{spall2022hybrid} or adaptive training (AT)~\cite{zhou2021large} (H-MBO)} are one type of the MBO methods that conduct forward pass in a real system and back-propagate the gradients with the aid of a physics-based simulator. 
\textit{Backpropagation through learned model (L-MBO)} is another method that gathers the real data from the optical computing system to train a neural proxy of the optical computing system~\cite{zheng2023dual, huo2023optical}. With such a learned proxy rather than the physics-based model utilized in H-MBO, the backward parameter update process can reduce the bias from the model mismatch between the simulator and the real system in L-MBO. During training, one has to optimize both the task's and proxy's parameters to optimize the task performance and improve the fidelity of the simulator. 
The task performance for the MBO, especially the L-MBO, is high while back-propagating with the digital twin also burdens the training process. 
The L-MBO methods implemented in optical computing are referred as \textit{hardware (camera)-in-the-loop methods}. These methods have recently been used in computational display~\cite{chakravarthula2020learned, peng2020neural} and camera ISP~\cite{tseng2022neural}. They are also known as \textit{real2sim} in photolithography for solving computational lithography tasks~\cite{zheng2023close, liao2022line}. 
Despite the computational burden introduced by the emulation process from the digital twin, another disadvantage of the MBO methods is that all the MBO approaches require images of input objects during training, adding further burden to the overall process.

\textbf{MFO methods.} Unlike the aforementioned MBO methods, MFO methods do not require constructing a digital twin of the optical computing system; they directly use the real system's output or intermediate measurements to update its parameters. \textit{Population-based algorithms} such as genetic algorithm (GA) methods~\cite{zhang2021efficient} use the output from the optical computing system to optimize parameters by simulating the process of natural selection. Still, the genetic algorithm's performance scales poorly to the dimensionality of parameters~\cite{grefenstette1986optimization}. 
\textit{Forward-forward training (FFT)} is a greedy multi-layer learning procedure for mortal computation~\cite{hinton2022forward} and it has been applied to train optical computing systems recently~\cite{momeni2023backpropagation, oguz2023forward}. However, labels must be added to each input sample during the training and testing. As a result, such training strategies require masking the input, which might be unrealistic in many scenarios, and the testing process of forward-forward training is prolonged.
In \textit{Trail and error}, perturbed parameters that enhance performance are retained, while those that do not are discarded, reverting to the previous state~\cite{bueno2018reinforcement, liu2022programmable}. This method, however, does not scale well with the dimensionality of parameters.
\textit{Intermediate measurement optical backpropagation (IMOB)} proposed using the adjoint variable method or direct feedback alignment, etc., to calculate gradients~\cite{Zhou:2020pr, nakajima2022physical, hughes2018training}. However, such methods require intermediate system measurements for every layer's output, which might be infeasible or require additional measurement setups. 

We propose the gradient-based MFO (G-MFO) training method using Monte Carlo gradient estimation to update the optical computing system parameters automatically. Our work shares similarities with Zhang \textit{et al.}, which uses the genetic algorithm for weight update~\cite{zhang2021efficient}, as our and their methods both generate a batch of sampled parameters during the training process. The difference is that our parameter updating process is more efficient owing to a score gradient estimator rather than their heuristic hill-climbing type algorithm. Moreover, our method does not need specific conditions, such as changing the system or manipulating the input object, which are more or less required in other MBO and MFO methods.

\subsection{Zeroth-Order Optimization and Its Applications to Computational Optics}

The zeroth-order (ZO) optimization uses finite differences from forward evaluations instead of the first-order path-wise gradient to calculate the gradient for updating the parameters of interest~\cite{mohamed2020monte, liu2020primer}. ZO has been utilized in many areas, such as differentiating through discontinuities in computer graphics~\cite{fischer2023zero}, optimizing contact dynamics in robotics~\cite{pang2023global, suh2022bundled}, and policy learning in reinforcement learning~\cite{williams1992simple}. ZO optimization has been implemented with both full-point estimation~\cite{kiefer1952stochastic}, and stochastic two-point estimation~\cite{spall1987stochastic}.
A significant advancement in ZO is to estimate the descent direction using multiple stochastic finite differences~\cite{spall1992multivariate} where Monte Carlo approximation is used in the process, also called evolutionary strategies~\cite{beyer2002evolution}. Moreover, techniques such as antithetic sampling, control variates~\cite{sanzalonso2024course}, or more advanced methods to reduce variance~\cite{kurenkov2021guiding, lehman2018more} in the Monte Carlo approximation process are used in multiple stochastic finite differences. Our work employs Monte Carlo gradient estimation~\cite{mohamed2020monte}, which resembles multiple stochastic finite differences.

\textit{Applications of ZO Methods to Computational Optics} have been relatively sparse compared to first-order methods. In the past, there have been demonstrations in computational optics tasks such as lens design~\cite{nagata2004lens}, auto-tuning structured light for depth estimation~\cite{chen2020auto}, and computer-generated holography~\cite{zhao2022model}. Using only forward evaluations for gradient updates makes it beneficial for optical computing-related tasks, such as what was mentioned as optical stochastic gradient descent (SGD) in the auto-tuning structured light work~\cite{chen2020auto}. The reason is that forward evaluation in optical computing is inexpensive while finding the first-order gradient is computationally costly and prone to errors. Our work highlights the advantages of using zeroth-order optimization for optimizing more complex systems, such as optical computing, and the efficacy of using Monte Carlo gradient estimation for efficient optimization.

\subsection{Application side of optical computing}

Optical computing, outperforming electrical computing in parallelism and processing speed, is pivotal for AI progress via optical neural networks (ONN) development~\cite{wetzstein2020inference}. ONNs encode inputs using light's spatial, temporal, and spectral characteristics to enable applications such as image classification, motion detection, and medical diagnosis~\cite{scalable_optical_learning_operator, liu2022programmable, reconfigurable_onn}. Furthermore, optical computing functions as an advanced neural network platform and directly processes optical signals, facilitating applications like phase reconstruction~\cite{mengu2022classification}, denoising~\cite{icsil2024all}, edge detection~\cite{park2022metasurface}, and unidirectional imaging~\cite{li2023unidirectional}. 

As a secondary contribution, our work showcases the application of optical computing for marker-free cell classification. Particularly relevant to our research is the study by Wang~\textit{et al.}~\cite{wang2023image}, which demonstrates the classification of fluorescent-labeled cells using an optical computing system. In contrast, our approach shows cell-classification based on their phase maps, instead of relying on fluorescent markers. This illustrates that optical computing can facilitate low-storage, high-speed, and marker-free cell analysis.

\section{Methodology} \label{sec: method}
In what follows, we mathematically detail the problem setup on training the optical computing system and our solution. We introduce problem formation in Subsec.~\ref{subsec: problem setup} and the conventional solution of using simulator-based training (SBT) in Subsec.~\ref{subsec: simulator-based design}. We then illustrate our solution to the problem, the gradient-based model-free optimization (G-MFO) for training the optical computing system in Subsec.~\ref{subsec: G-MFO methodology}. Finally, we illustrate the optical computing system and the related simulators in Subsec.~\ref{subsec: system description}, which we will use to demonstrate the performance of our method.

\subsection{Problem setup} \label{subsec: problem setup}
 We are interested in learning the optimal weight $w\in\mathbb{R}^{H}$ for the optical computing system on a desired task with a training dataset $\mathcal{D} = \{x_i, y_i\}_{i=1}^N$, where $N$ is the size of the dataset, $H$ is the number of trainable parameters in $w$, and $x$ and $y$ denote the input and target of interest, respectively. A function $f_{sys}(\cdot,w)$ maps $x\rightarrow y$ through this optical computing system with $w$. Specifically, in the image classification task based on the diffractive optical computing system we work on, $f_{sys}$ denotes the optical mapping from the input image $x$ to the output label $y$, and $w$ is the optical computing weight in the form of phase value modulation. 
 
During training, one can minimize the cost function $J(w)$ as the mean task-specific loss across the entire training data set $\mathcal{D}$
:

\begin{subequations}
\begin{align}
          \operatorname*{arg\,min}_w J(w): &=\mathbb{E}[\mathcal{L}(\mathcal{D}, w)], \\
                    &= \frac{1}{N}\sum_{i=1}^{N}\mathcal{L}(f_{sys}(x_i, w), y_i), \label{eq:objective function}
\end{align}
\end{subequations}
where $\mathcal{L}$ is the task-specific loss function, we use cross-entropy loss~\cite{brier1950verification} since we deal with image classification tasks throughout this paper. 

We use gradient descent-based search to find the optimal $w$ to minimize the objective function $J(w)$:
\begin{equation} \label{eq:grad descent}
    w = w - \alpha \nabla_{w} J(w),
\end{equation}
where $\nabla_{w}$ represents the gradient operator that collects all the partial derivatives of a function concerning parameters in $w$, and $\alpha$ is the learning rate. 

It is straightforward to use the backpropagation method~\cite{rumelhart1986learning} to take the gradient through $f_{sys}$ and finds the gradient $\nabla_{w} J(w)$ as:
\begin{equation}
    \nabla_w J(w)= \frac{1}{N}\sum_{i=1}^N \nabla_w \mathcal{L}({f_{sys}(x_i, w)}, y_i),
    \label{eq: FOBG}
\end{equation}
when we have an accurate and differentiable $f_{sys}$ modeling.
However, this is the case for training digital neural networks~\cite{rumelhart1986learning}, but not when we train a real-world optical computing system. \textbf{Thus, this paper's critical aim is finding an accurate gradient estimation of $\nabla_w J(w)$ to update optical computing weight $w$ in a real-world optical system}.

\begin{figure}[h!]
\centering
\includegraphics[width=.9\linewidth]{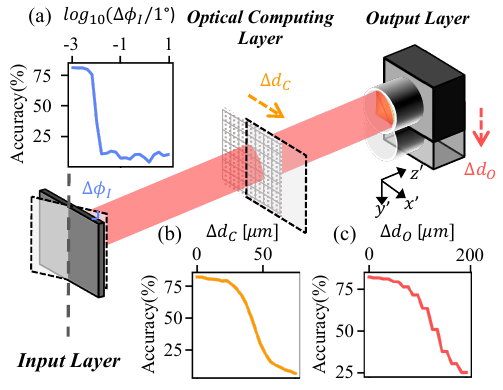}

\caption{\textbf{System misalignment in a real optical computing system degenerates the performance of the optical computing system trained solely with a physics-based simulator.} 
(a) Testing accuracy drops to $36.4\%$ from $82.2\%$ when having a misaligned rotation angle $\Delta \phi_I=0.01\degree$. 
(b) Testing accuracy drops to $51.0\%$ from $82.2\%$ when the x'-axis misalignment of the optical computing layer $\Delta d_c$ is $41.1 \: \mu m$. 
(c) Testing accuracy decreases to $50.9\%$ from $82.2\%$ when the y'-axis misalignment of the output layer $\Delta d_o$ is $62.4 \: \mu m$.
}
\verb||\label{fig: misalignment degrades the performance}
\vspace{-0.4cm}
\end{figure}

\subsection{In silico simulator-based training (SBT)} \label{subsec: simulator-based design}

\textbf{Back-propagation through the simulator $\hat{f}_{sys}$.} In a real-world optical computing system, since we do not have an exact functional expression of $f_{sys}$, the \textbf{simulator-based training} (SBT) builds a simulator $\hat{f}_{sys}$ as the differentiable approximation of $f_{sys}$ (Fig.~\ref{fig:mfo}b). 
A naive training strategy is substituting $f_{sys}$ in Eq.~\ref{eq: FOBG} with the simulator $\hat{f}_{sys}$ and applying in silico training on the simulator:

\begin{equation}
    \nabla_w J(w)= \frac{1}{N}\sum_{i=1}^N \nabla_w \mathcal{L}({\hat{f}_{sys}(x_i, w)}, y_i).
    \label{eq: sbt}
\end{equation}
The resulting $\nabla_w J(w)$ is used in Eq.~\ref{eq:grad descent} to update $w$.
After training, the optimized $w$ is uploaded to the real optical computing system $f_{sys}$ to test the performance.

\textbf{Simulation-to-reality gap.} The aforementioned simulator-based training is based on backpropagation through the simulator $\hat{f}_{sys}$. The "sim2real" gap is low (i.e., the gradient $\nabla_w J(w)$ is an accurate estimation) when the simulator $\hat{f}_{sys}$ is similar to $f_{sys}$. However, this assumption does not hold in many prototypes of optical computing systems where inadequate modeling and misalignment between the optical elements decay the performance of the SBT during the "sim2real" transfer. We use a simulator described in Subsec.~\ref{subsec: system description} to assess the adverse effect of misalignment on the image classification results. We impose various misalignments to a well-trained ideal optical computing system and measure drops in classification accuracy. For instance, we show in Fig.~\ref{fig: misalignment degrades the performance} that slightly laterally misaligning the optical computing layer by $41.1 \: \mu m$ reduces the classification accuracy by $31.2\%$.

\subsection{In situ G-MFO}\label{subsec: G-MFO methodology}
Our solution for solving the aforementioned "sim2real" gap issue in Subsec.~\ref{subsec: simulator-based design} is in situ learning the optical computing weight $w$ with the gradient-based model-free optimization (G-MFO). In situ learning enables us to access the output of $f_{sys}$ and is feasible on the hardware side because we can use programmable devices such as spatial light modulators to update $w$. The challenging part is designing a training strategy that efficiently uses the output of actual system $f_{sys}$ to construct an unbiased gradient estimator. Here, we use the score gradient estimator to calculate the gradient~\cite{schulman2015gradient, mohamed2020monte} for the backward update of parameters in $w$ while circumventing the construction of $\hat{f}_{sys}$, a biased and resource-intensive numerical modeling of $f_{sys}$. 

\textbf{Back-propagation through the weights distribution $p$.} In our score gradient estimator for model-free optimization, we model optical computing weight $w$ as a random variable that follows a 
probability distribution: $w\sim p(w|\theta)$ and rewrite the cost function $J(w)$ in  Eq.~\ref{eq:objective function} as a probability cost function $J(\theta)$:
\begin{subequations}
\begin{align}
        \operatorname*{arg\,min}_{\theta} J(\theta): 
          &= \frac{1}{N}\sum_{i=1}^{N} \mathbb{E}_{p(w|\theta)}[\mathcal{L}(f_{sys}(x_i, w), y_i)],\\
          &= \frac{1}{N}\sum_{i=1}^{N} \int {p(w|\theta)}\mathcal{L}(f_{sys}(x_i, w), y_i) \diff w,
\end{align}\label{eq: prob obj}
\end{subequations}
where the probability distribution $p(w|\theta)$ is continuous in its domain and differentiable concerning its distribution parameter $\theta$. 
Accordingly, the original goal of optimizing $w$ is reformulated as finding a most likely distribution $p(w|\theta)$ that minimizes the cost function $J(\theta)$ in Eq.~\ref{eq: prob obj}. 
Specifically, we model the distribution $p$ as a multivariate normal distribution with $\theta = \{\mu, \sigma^2\}$ and $p(w|\theta) = \mathcal{N}(w;\mu, \sigma^2)$ for optimizing the continuous phase-valued weight to be uploaded onto the SLM in our work.

To update distribution parameter $\theta$ with the gradient descent Eq.~\ref{eq:grad descent}, we take the gradient of the cost function $J(\theta)$ in Eq.~\ref{eq: prob obj}:
\begin{subequations}
\begin{align}
        \nabla_{\theta} J(\theta) 
        &= \nabla_{\theta} \frac{1}{N}\sum_{i=1}^{N} \int {p(w|\theta)}\mathcal{L}(f_{sys}(x_i, w), y_i) \diff w ,\\
        &= \frac{1}{N}\sum_{i=1}^{N} \int \mathcal{L}(f_{sys}(x_i, w), y_i) \nabla_{\theta}{p(w|\theta)} \diff w .
\end{align}
\end{subequations}
Applying the log derivative trick, we have:
\begin{equation}
        \nabla_{\theta} J(\theta) 
        = \frac{1}{N}\sum_{i=1}^{N} \int p(w|\theta) \mathcal{L}(f_{sys}(x_i, w), y_i) \nabla_{\theta} \log{p(w|\theta)} \diff w .\label{eq: prob obj int}
\end{equation}
Then we apply Monte Carlo integration to approximate the integral value in Eq.~\ref{eq: prob obj int} by first drawing $M$ independent samples $\{w_j\}_{j=1}^M$ from the distribution $p(w|\theta)$ and then computing the average function value evaluated in these samples:
        \begin{subequations}
        \begin{align}
        \nabla_{\theta} J(\theta) 
        &=\frac{1}{N}\sum_{i=1}^{N} \frac{1}{M}\sum_{j=1}^{M} \mathcal{L}(f_{sys}(x_i, w_j), y_i) \nabla_{\theta} \log{p(w_j|\theta)} \label{eq: prob obj monte carlo int},\\
        &=\frac{1}{M}\sum_{j=1}^{M} [\frac{1}{N}\sum_{i=1}^{N}  \mathcal{L}(f_{sys}(x_i, w_j), y_i)] \nabla_{\theta} \log{{p(w_j|\theta)}},\\
        &=\frac{1}{M}\sum_{j=1}^{M} r(w_j) \nabla_{\theta} \log{{p(w_j|\theta)}}. 
        \label{eq: prob grad}
\end{align}
\end{subequations}
Here we define $r(w_j)=\frac{1}{N}\sum_{i=1}^{N} \mathcal{L}(f_{sys}(x_i, w_j), y_i)$ as the negative reward corresponding to each weight $w_j$. We use Equation~\ref{eq: prob grad} as the score gradient estimator for G-MFO, 
where the score function is $\nabla_{\theta} \log{p(w_j|\theta)}$, which has been widely used in other areas, such as policy gradient algorithms in reinforcement learning~\cite{williams1992simple} and diffusion models~\cite{song2019generative}.

\textit{Variance reduction.} The main risk of using the score gradient estimator is the high variance that comes from the Monte Carlo integration step that transits Eq.~\ref{eq: prob obj int} to Eq.~\ref{eq: prob obj monte carlo int}. Such a sampling-based integration step has high variance because different sets of random samples may lead to significantly different integral estimates. We reduce the variance by subtracting the $r(w_j)$ with baseline value $\bar{r} = \frac{1}{M} \sum_{j=1}^M r(w_j)$ while keeping the bias of gradient unchanged~\cite{mei2023role}:
\begin{equation}
    \nabla_{\theta} J(\theta) = 
    \frac{1}{M}\sum_{j=1}^{M} (r(w_j)- \bar{r}) \nabla_{\theta} \log{{p(w_j|\theta)}}. \label{eq: prob grad simplified}
\end{equation}

\begin{figure}[h!]
\centering
\includegraphics[width=0.98\linewidth]{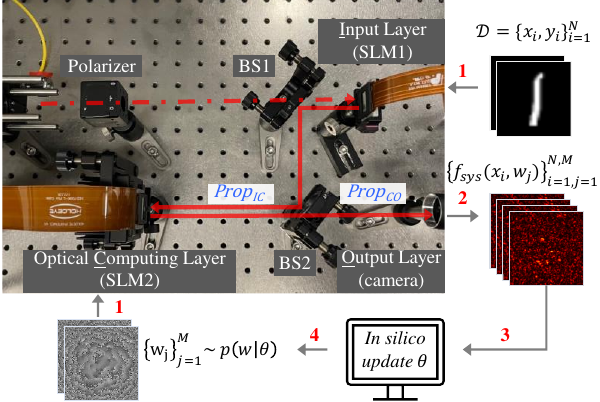}
\caption{\textbf{A visual illustration of the training process for the G-MFO experiment on the real system.} }
\label{fig: MFO_workflow}
\end{figure}

\textbf{Training recipe of G-MFO.} 
We perform the following steps in each training epoch (see Fig.~\ref{fig: MFO_workflow}): 1. Sample a group of phase-valued optical computing weights $\{w_j\}_{j=1}^M$ from the distribution $w_j \sim p(w|\theta)$ and upload them to the optical computing layer. Upload classification data $\mathcal{D}=\{x_i, y_i\}_{i=1}^N$ to the input layer; 2. Collect the optical computing system's output $\{f_{sys}(x_i, w_j)\}_{i=1, j=1}^{N, M}$ for each pair of input and weight; 3. Update the distribution parameter $\theta$ in silico using Eqs.~\ref{eq:grad descent} and~\ref{eq: prob grad simplified}; 4. Sample a new group of $\{w_j\}_{j=1}^M$ from the updated distribution $w_j \sim p(w|\theta)$, which is ready to be uploaded to the optical computing layer in the next epoch.
The algorithm iterates these steps until convergence. 

After minimizing the objective function Eq.~\ref{eq: prob obj}, we export $w_j$ with the smallest $r(w_j)$ as the output weight $w^{\star}$. This performs better than setting $w^{\star}$ as the sampled mean $\mu$. 
An algorithmic overview of the above training recipe is shown in Algorithm~\ref{alg:mfo}.

\begin{algorithm}
\caption{Algorithmic overview of G-MFO.}\label{alg:mfo}
\begin{algorithmic}[1]
\State \textbf{Input: }  {Classification dataset $\mathcal{D}=\{x_i, y_i\}_{i=1}^N$, learning rate $\alpha$, number of sampled weights $M$, optical computing system $f_{sys}$, distribution parameter $\theta=\{\mu, \sigma^2\}$, loss function $\mathcal{L}$, epochs $K$.}
\State \textbf{Output: }{Optimized optical computing weight $w^{\star}$.}

\For{$k$ in range $K$}
    \State{Sample $\{w_j\}_{j=1}^M$ from distribution $p(w|\theta)$.}
    \State{\emph{$\triangleright$ in situ evaluate $\{w_j\}_{j=1}^M$.}}
    \For{$j$ in range $M$}\\
        \State{$r(w_j)\gets \frac{1}{N}\sum_{i=1}^{N} \mathcal{L}(f_{sys}(x_i, w_j), y_i).$}
    \EndFor
    \State{\emph{$\triangleright$ in silico update $\theta$.}}
    \State{Calculate $\nabla_{\theta} J(\theta)$ via Eq.~\ref{eq: prob grad simplified}.}
    \State{$\theta \gets \theta -\alpha\nabla_{\theta} J(\theta).$ }
\EndFor
\State $w^{\star}\gets w_j$ with the smallest $r(w_j)$.
\end{algorithmic}
\end{algorithm}

\subsection{Experiment and Simulation Details of Our Diffractive Optical Computing System} \label{subsec: system description}

We validate the effectiveness of the proposed G-MFO-based training strategy on a single-layer optical computing system and a two-layer diffractive optical computing system.
Our work focuses on the training strategy rather than the design or adavantage of a specific optical computing system. Therefore, we do not consider nonlinearity in constructing our optical computing system, which has recently been implemented using saturation effects~\cite{zhou2021large, wang2023image, zhang2024broadband}. The validity of using a multi-layer versus a single-layer system without nonlinearity has been analyzed empirically in~\cite{kulce2021all}, where the authors attribute the improvement to increased energy resulting from the multi-layer settings. Details on the experimental setups ($f_{sys-1-layer}$, $f_{sys-2-layer}$) and the corresponding simulators ($\hat{f}_{sys-1-layer}$, $\hat{f}_{sys-2-layer}$) are in the Supplement Sec. S.1.

\begin{figure*}[t!]
\newcolumntype{P}[1]{>{\centering\arraybackslash}p{#1}}
\centering
\includegraphics[width=0.95\linewidth]{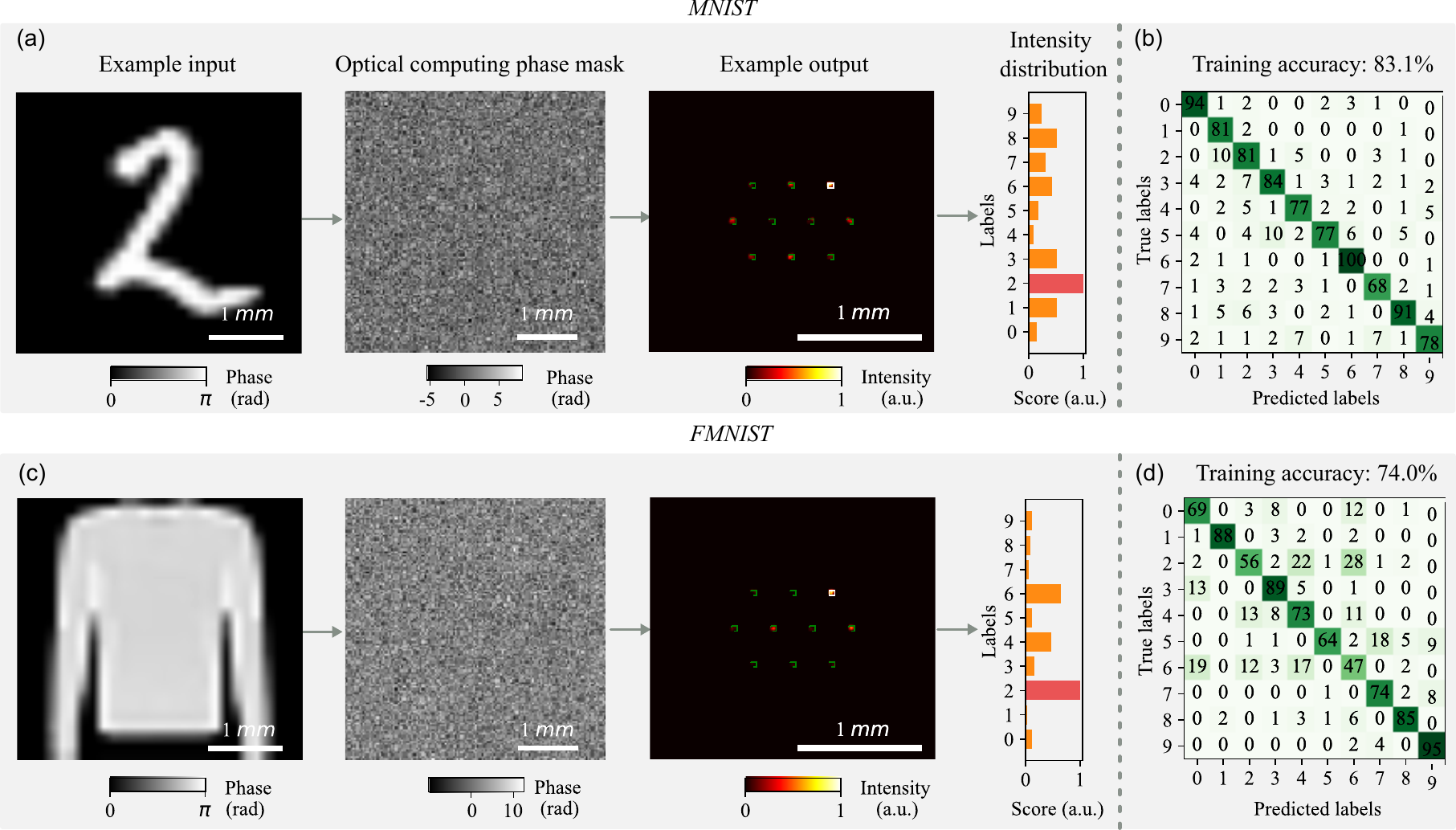}
\caption{\textbf{Visualization of the single-layer optical computing system's experimental outputs and confusion matrices trained with G-MFO.} (a) An input phase object digit '$2$' from the MNIST dataset is modulated by the optical computing layer with weight $w$ trained using G-MFO. 
The system correctly predicts the input as digit ‘$2$’, as the output image has the largest intensity at the region corresponding to digit '$2$'. (b) Confusion matrix on the MNIST dataset with a training accuracy of $83.1\%$. (c) An example of a 'pullover' from the FMNIST dataset is correctly predicted. (d) Confusion matrix on the FMNIST dataset with a training accuracy of $74.0\%$.
}
\verb||\label{fig: G-MFO result}
\end{figure*}

\vspace{-0.2cm}
\section{Results}\label{sec: results}

We aim to answer the following research questions in this section:

\begin{itemize}
    \item How does G-MFO perform in the numerical simulations and experiments?
    \item What are the advantages and limitations of G-MFO?
    \item How is G-MFO's applicability on the marker-free cell classification task?
\end{itemize}

Thus, we numerically and experimentally evaluate the general performance of our G-MFO method on the open-source MNIST and FMNIST datasets (Subsec.~\ref{subsec: general performance}). 
We then illustrate G-MFO's advantage of memory- and computation-efficient training in Subsec.~\ref{subsec: source efficient training by G-MFO.}. Moreover, we analyze the limitation of our method in Subsec~\ref{subsec: G-MFO exhibits curse of dimensionality.}. Lastly, we demonstrate the G-MFO method on a novel application of marker-free classifying white blood cells in Subsec.~\ref{subsec: classify WBC.}.

{We compare our method with the following baselines:}

\begin{itemize}
    \item \textbf{Ideal:} The in silico simulation results without introducing any artificial misalignment during the simulation. This corresponds to the best achievable result for an optical computing system. 
    \item \textbf{SBT:} Simulator-based training with digitally aligned simulator. 
    \item \textbf{HBT:} The hybrid training strategies with a physics-based simulator.
    \item \textbf{L-MBO:}  We pre-train a proxy using the task dataset to emulate the real system and then optimize the optical computing task based on the pre-trained proxy.
    \item \textbf{Zeroth-order optimization baselines:} Including two-point estimation on a random direction~\cite{spall1987stochastic} and full-point estimation~\cite{kiefer1952stochastic, chen2020auto}.
\end{itemize}
The training details for simulations and experiments are in Supplement Sec. S.3.
Supplement Sec. S.6 provides a detailed description of our implementation of the HBT baseline. Supplement Sec. S.7 discusses and compares L-MBO.

\subsection{General performance evaluation on the MNIST and FMNIST dataset}~\label{subsec: general performance}

\vspace{-0.8cm}
\rev{\subsubsection{Simulation comparison between G-MFO and other zeroth-order optimization methods in simulation on a small dataset}\label{subsec: simulation mfo comparison}}

\rev{We use a single-layer optical computing simulator to compare our method with two zeroth-order (ZO) baselines: the full-point ZO method~\cite{chen2020auto, kiefer1952stochastic} and the two-point ZO method~\cite{spall1987stochastic}. This comparison mainly evaluates the training performance, including accuracy, execution, and convergence speed. Therefore, we use a small dataset of 100 training, 100 validation, and 100 testing samples.}

\begin{figure}[h!]
\centering
\includegraphics[width=0.98\linewidth]{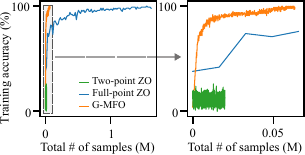}
\caption{\rev{\textbf{G-MFO balances sample efficiency and task performance for the optimization process.} We compare the training curves of the two-point zeroth-order (ZO) method (green), the full-point ZO method (blue), and our G-MFO (orange) method. The full-point ZO method and G-MFO achieve $99\%$ training accuracy on the small dataset. Notably, the total number of samples of G-MFO is only approximately $4\%$ of that required by the full-point ZO method to achieve an accuracy of $99\%$. Additionally, the two-point ZO method fails to optimize effectively. The right panel presents an enlarged view of the selected region from the left panel.}}
\label{fig: 0order comparison}
\end{figure}

\rev{We present the simulation results and training time per epoch in Table~\ref{tab: 0th order comparison result}, and the curves of training accuracy in Fig.~\ref{fig: 0order comparison}. The results show that sufficient samples are necessary for successful optimization. Specifically, the two-point method, with two samples per epoch, only achieves a training accuracy of $22\%$, whereas both the full-point method, using $128 \times 128$ samples per epoch, and the G-MFO method, using $128$ samples per epoch, achieve a training accuracy of $99\%$. Additionally, G-MFO has improved sample efficiency than the full-point optimization method, requiring only $4\%$ of the samples to achieve an accuracy of $99\%$ compared to the full-point optimization method.}

\begin{table}[h!]
\newcolumntype {C}[1]{>{\centering\arraybackslash}m{#1}}
\arrayrulecolor{black}
\centering
\renewcommand{\arraystretch}{1.5}
\begin{tabular}{C{0.16\linewidth} | C{0.08\linewidth} | C{0.08\linewidth}| C{0.08\linewidth} |C{0.15\linewidth}| C{0.15\linewidth}}
\hline
             &\multicolumn{3}{c|}{Accuracy}    &\# of samples per epoch  &Training time per epoch\\\hline
             & Train             & Val                 & Test \\\hline\hline
Full-point~\cite{chen2020auto, kiefer1952stochastic}         & $99.0\%$          & $75.0\%$            & $59.0\%$   & $128\times128$& $113~min$\\
Two-point~\cite{spall1987stochastic}         & $22.0\%$          & $27.0\%$            & $26.0\%$   & 2             & $2~s$ \\

G-MFO (Ours) & $99.0\%$          & $73.0\%$            & $60.0\%$   & $128$           & $2~min$\\
\hline
\end{tabular}
\caption{\rev{\textbf{Numerical comparison of G-MFO and other zeroth-order (ZO) optimization methods on FMNIST dataset on a simulated single-layer diffractive optical computing system. 
} We show results from the full-point ZO method, the two-point ZO method, and G-MFO. }}
\label{tab: 0th order comparison result}
\end{table}

\begin{table}[h!]
\newcolumntype {C}[1]{>{\centering\arraybackslash}m{#1}}
\arrayrulecolor{black}
\centering
\renewcommand{\arraystretch}{1.5}
\begin{tabular}{C{0.16\linewidth} | C{0.072\linewidth} | C{0.072\linewidth}| C{0.072\linewidth}| C{0.072\linewidth}| C{0.072\linewidth}| C{0.072\linewidth}}
\hline
             &\multicolumn{3}{c|}{MNIST}          & \multicolumn{3}{c}{FMNIST} \\ \cline{2-7}
             & Train      & Val      & Test       & Train         & Val         & Test \\\hline\hline
\textcolor{gray}{Ideal}       & \textcolor{gray}{$90.0\%$}          & \textcolor{gray}{$89.3\%$}            & \textcolor{gray}{$89.5\%$}            & \textcolor{gray}{$82.2\%$}         & \textcolor{gray}{$81.2\%$}           & \textcolor{gray}{$80.5\%$} \\
\hline
HBT          & $78.0\%$         & $78.7\%$           & $78.4\%$         & $61.6\%$          & $61.4\%$           & $59.8\%$\\
G-MFO (Ours)         & $\textbf{86.2\%}$         & $\textbf{86.2\%}$           &$\textbf{87.0\%}$         & $\textbf{75.4\%}$          & $\textbf{75.4\%}$           & $\textbf{74.1\%}$ \\ 

\hline
\end{tabular}
\caption{\textbf{Numerical performance comparison on MNIST and FMNIST datasets on two-layer diffractive optical computing system.
}}
\label{tab: two layer simulation result}
\end{table}

\vspace{0.3cm}
\subsubsection{Simulation results on the two-layer optical computing system $\hat{f}_{sys-2-layer}$.}\label{subsec: simulation results two layer}

We evaluate our method's accuracy by conducting performance tests on a two-layer optical computing simulator $\hat{f}_{sys-2-layer}$ utilizing two classical image classification datasets: MNIST~\cite{deng2012mnist} and FMNIST~\cite{xiao2017fashion}. Each optical computing layer contains $128 \times 128 = 16,384$ trainable parameter, and the system contains $128 \times 128 \times 2 = 32,768$ parameters. Furthermore, to assess the robustness of the G-MFO method against system misalignment, we intentionally introduce a slight misalignment on the positive x'-axis direction, amounting to $40\mu m$ (equivalent to 5 pixels on SLM1) on the optical computing layer1, and $18.7\mu m$ (corresponding to 5 pixels on SLM2), on the optical computing layer2.
We use $10,000$ data to train, $10,000$ data to validate, and $10,000$ data to test.

The simulation outcomes for the MNIST and FMNIST datasets are in Tab.~\ref{tab: two layer simulation result}. 
The ideal testing accuracy reaches $89.5\%$ on MNIST and $80.5\%$ on FMNIST. However, the presence of misalignment reduces the accuracy to $78.4\%$ and $59.8\%$ using HBT, a decrease of approximately $10\%$ and $20\%$. In contrast, the G-MFO method hits an accuracy of $87.0\%$ and $74.1\%$, effectively mitigating the detrimental impact of system misalignment.

\begin{table}[h!]
\newcolumntype {C}[1]{>{\centering\arraybackslash}m{#1}}
\arrayrulecolor{black}
\centering
\renewcommand{\arraystretch}{1.5}
\begin{tabular}{C{0.16\linewidth} | C{0.072\linewidth} | C{0.072\linewidth}| C{0.072\linewidth}| C{0.072\linewidth}| C{0.072\linewidth}| C{0.072\linewidth}}
\hline
             &\multicolumn{3}{c|}{MNIST}          & \multicolumn{3}{c}{FMNIST} \\ \cline{2-7}
             & Train      & Val      & Test       & Train         & Val         & Test \\\hline\hline
\textcolor{gray}{Ideal}        & \textcolor{gray}{$92.7\%$}          & \textcolor{gray}{$84.3\%$}            & \textcolor{gray}{$82.2\%$}            & \textcolor{gray}{$85.6\%$}         & \textcolor{gray}{$79.9\%$}           & \textcolor{gray}{$76.4\%$} \\
\hline
\hline
SBT          & $81.9\%$         & $74.4\%$           & $69.3\%$         & $68.3\%$          & $64.1\%$           & $60.9\%$ \\
HBT          & $81.9\%$         & $75.5\%$           & $72.8\%$         & $68.3\%$          & $68.7\%$           & $65.8\%$ \\
G-MFO (Ours)         & $\textbf{83.1\%}$         & $\textbf{77.8\%}$           &$\textbf{73.6\%}$         & $\textbf{74.0\%}$          & $\textbf{71.1\%}$           & $\textbf{70.4\%}$ \\ 
HBT+G-MFO (Ours)      & $\textbf{87.0\%}$& $\textbf{80.2\%}$  & $\textbf{77.7\%}$ &$\textbf{75.0\%}$  & $\textbf{71.7\%}$   & $\textbf{70.2\%}$  \\
\hline
\end{tabular}
\caption{\textbf{Experimental performance comparison on MNIST and FMNIST datasets on the single-layer optical computing system.} Results of the ideal mode are from the simulator, whose parameters are determined from experiments while we impose no misalignment on the simulator. The lower four rows are experimental results from SBT, HBT, G-MFO, {and HBT+G-MFO (G-MFO fine-tuning upon the result from HBT)}. 
Our method outperforms the SBT and HBT methods on the MNIST and FMNIST datasets.
} 
\label{tab: MNIST and FMNIST result}
\vspace{-0.2cm}
\end{table}

\subsubsection{G-MFO outperforms hybrid training (HBT) experimentally on a single-layer optical computing system $f_{sys-1-layer}$.}\label{subsec: experiment result one layer}

We conduct experiments on a real single-layer optical computing system $f_{sys-1-layer}$ on the MNIST and FMNIST datasets. We include in silico SBT and in situ HBT methods utilizing digitally aligned simulators as the comparison baselines. The training, validation, and testing phases each utilize a dataset of 1,000 samples. 

Tab.~\ref{tab: MNIST and FMNIST result} quantitatively shows that our method achieves higher classification accuracy than the HBT and SBT methods on both datasets in the experiments. The SBT method performs poorly due to the reality gap between the simulator and the real system. 
The HBT method suffers from the bias between $\hat{f}_{sys}$ and ${f}_{sys}$ in the backward process, while the G-MFO bypasses the bias-sensitive modeling and updates gradients solely with $f_{sys}$. 
We further show G-MFO's capability to fine-tune the HBT result. The last row of Tab.~\ref{tab: MNIST and FMNIST result} shows that the unbiased G-MFO method further improves the results of HBT. Moreover, we also empirically find that fine-tuning outperforms G-MFO only. Fig.~\ref{fig: G-MFO result} visualizes some experimental outputs and confusion matrices using the G-MFO method.

\subsubsection{Experimental results on the two-layer optical computing system $f_{sys-2-layer}$.}\label{subsec: experiment result two layer}

We also perform experiments on training a real two-layer optical computing system $f_{sys-2-layer}$ on the MNIST dataset. For this experiment, we utilize a dataset comprising 10,000 instances for training, another 10,000 for validation, and a separate set of 10,000 for testing. The training, validation, and testing accuracy are  $80.43\%$, $81.1\%$, and $80.3\%$, respectively. 

Compared with the simulated G-MFO outcomes presented in Tab.~\ref{tab: two layer simulation result}, the experimental accuracy is lower. We hypothesize that this discrepancy is attributed, in part, to mechanical disturbances encountered during the multi-day span of the experiment. Moreover, the training accuracy from a two-layer system is lower than that of a single-layer system, while the validation and testing accuracy is higher. This is because we use more samples when training a two-layer system, alleviating overfitting.

\begin{figure*}[h]
\newcolumntype{P}[1]{>{\centering\arraybackslash}p{#1}}
\centering
 \includegraphics[width=0.9\linewidth]{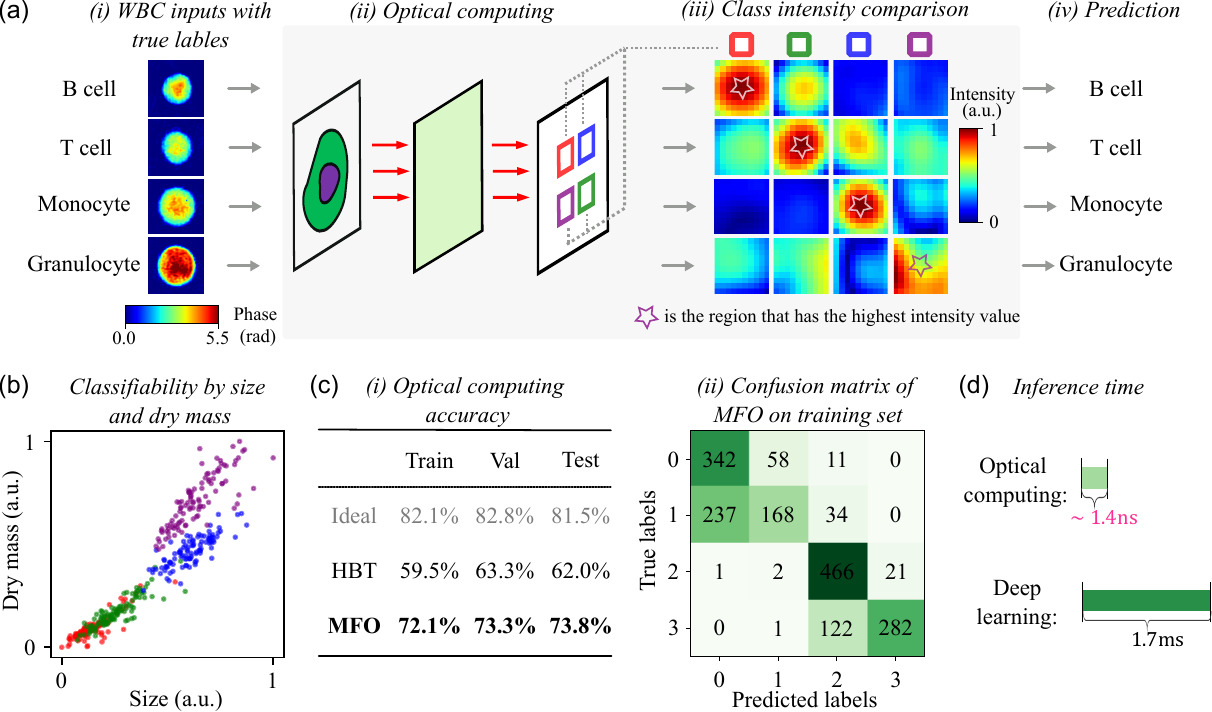}
\caption{\textbf{All optical marker-free cell classification enabled by G-MFO training.} (a) Our trained optical computing system classifies four WBC subtypes, including B cell (red), T cell (green), monocyte (blue), and granulocyte (purple), in a marker-free manner. The system uses the (i) marker-free phase information of the WBCs to (ii) perform optical classification. The output plane of the system has different regions corresponding to different classes. We (iii) measure the intensity values in these regions and (iv) choose the class with the highest intensity as the prediction. (b) shows that non-morphological features, including size and dry mass, cannot divide the four subtypes well. In contrast, (c)(i) shows the optical computing system trained with our G-MFO method can classify the WBC subtypes with a training/validation/testing accuracy of $72.1\%/73.3\%/73.8\%$, which is higher than that of the HBT method. (c)(ii) shows the confusion matrix of G-MFO results on the training set. (d) The optical computing system produces its output at the speed of light ($1.4 \: ns$). This is faster than the electronic computing with ResNet10~\cite{wbcphase2021shu, he2016resnet}, which has an inference time of $1.7 \: ms$.
}
\verb||\label{fig: all-optical cell classfier}
\end{figure*}

\subsection{Advantage: memory- and computation-efficient training enabled by G-MFO}\label{subsec: source efficient training by G-MFO.}

\begin{figure}[h!]
\centering
\includegraphics[width=0.8\linewidth]{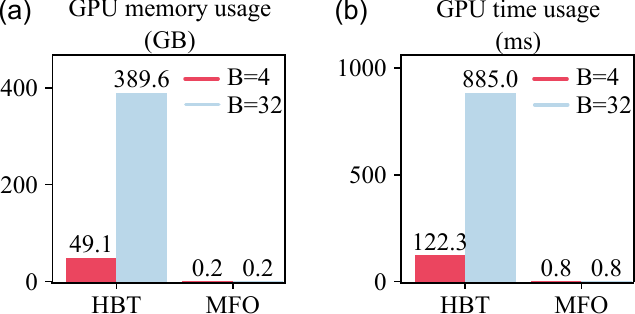}
\caption{\textbf{Advantage: G-MFO saves in silico computing resources during training.} 
We compare the GPU (a) memory and (b) time consumption of HBT and G-MFO methods for training one batch of inputs with batch sizes of $B=4$ and $B=32$. G-MFO is more efficient than HBT during the in situ training process in memory and time. Moreover, G-MFO's resource consumption does not increase as the batch size increases, unlike HBT.
}  
\verb||\label{fig: resource consuming}
\end{figure}

In addition to the predicting accuracy discussed in the previous subsection, our G-MFO method has an advantage over other training algorithms regarding GPU time and memory efficiency. 
The SBT and HBT methods compute $\hat{f}_{sys}(x, w)$ and $\nabla_{w}\hat{f}_{sys}$ (Fig.~\ref{fig:mfo}b) for each input $x$ in silico, which requires a lot of in silico computation resources. 
In contrast, our G-MFO method executes the calculation of $f_{sys}(x, w)$ in the light-speed real optical computing system, but not in the computation- and memory-heavy simulator $\hat{f}_{sys}(x,w)$.
The only step of our G-MFO method that consumes in silico computational resources is the one described in Eq.~\ref{eq: prob grad simplified}, where we calculate the score gradient. The in silico computational resources consumption in this step is low because it only scales with the dimension of $w$ and has no relation to the complexity of the system's light transport $\hat{f}_{sys}$.

We compare our G-MFO method with HBT in GPU memory and time usage during experiments on a single-layer optical computing system in Fig.~\ref{fig: resource consuming}. Our G-MFO method requires far less GPU time and memory than the HBT method during the training.

\subsection{Limitation: G-MFO exhibits the curse of dimensionality}\label{subsec: G-MFO exhibits curse of dimensionality.}
Our method is not without its limits. Our G-MFO training relies on Monte Carlo integration. It thus inherits the \textit{curse of dimensionality} from the Monte Carlo integration~\cite{bellman1959adaptive}. That is, the number of samples $M$ needed to estimate the integration in Eq.~\ref{eq: prob obj int} with a given level of accuracy grows exponentially concerning the $H$, the number of trainable parameters (i.e., dimensionality) of the function.
This is also discussed and alleviated with \textit{variance reduction} in the previous Sec.~\ref{subsec: G-MFO methodology}. However, the G-MFO strategy presented in
this paper is still sample-inefficient, though unbiased and memory-efficient. 
We need to either limit the number of trainable parameters $H$, or sample a large number of varied optical computing weights $\{w\}^M_{j=1}$ from distribution $p(w|\theta)$ in every iteration to make G-MFO's gradient less noisy. 
The former limits our method's design space, while the latter requires more executions on the real system, which prolongs the training time. 

\begin{figure}[h!]
\centering
\includegraphics[width=0.95\linewidth]{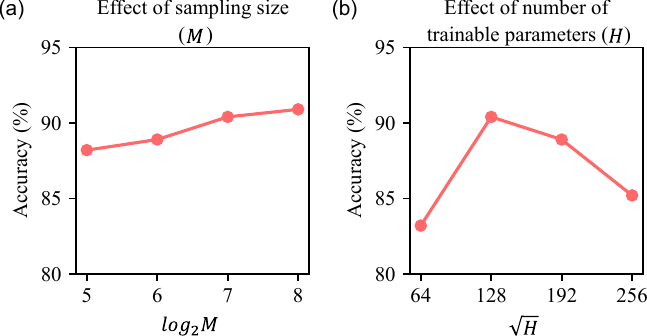}
 \caption{\textbf{Limitation: G-MFO exhibits the curse of dimensionality.} (a) When the sampling size of optical computing weights $M$ rises from $2^5=32$ to $2^7=128$, the training accuracy increases by $2.2\%$. However, increasing $M$ from $2^7=128$ to $2^9=256$ only improves $0.5\%$. Here, we fix the $H$ as $128^2$. (b) G-MFO achieves the highest accuracy when $H = 128^2$ under a sampling size $M=128$ while further enlarging $H$ fails the result.
 }  
\verb||
\label{fig: curse of dimensionality}
\end{figure}

We quantitatively investigate the influence of this limitation in Fig.~\ref{fig: curse of dimensionality} through simulations on a one-layer optical computing system $\hat{f}_{sys-1-layer}$, employing $1,000$ training samples from the MNIST dataset. The figure demonstrates how $M$ and $H$ impact the training performance of G-MFO.
\vspace{-0.1cm}
Shown in Fig.~\ref{fig: curse of dimensionality}(a), G-MFO requires a $M>=128$ to achieve a training accuracy of $>90\%$ given $H=128^2$. 
Moreover, in Fig.~\ref{fig: curse of dimensionality}(b), our G-MFO method fails when increasing $H$ beyond $128^2$ while keeping the sampling size $M$ fixed to $128$.

\subsection{Application: all-optical classification on marker-free cellular dataset}\label{subsec: classify WBC.}

For the first time, we demonstrate the capability of an optical computing system for marker-free cell analysis, trained by our G-MFO algorithm (Fig.~\ref{fig: all-optical cell classfier}). We work on the white blood cells (WBC), whose abnormal subtype percentages indicate the immune system’s malfunction or infectious disease \cite{wbcfunction1, wbcfunction2}. We include details of the WBC phase map dataset in Supplement Sec. S.2. Previously, researchers used machine learning methods to classify WBC subtypes, including monocyte, granulocyte, B cell, and T cell, by their morphology in a marker-free way~\cite{wbcbrightfield2019Nassae,wbcphase2021shu}. However, the analysis process is computationally heavy and time-consuming. 

Here, we accelerate the marker-free cell analysis process via computing with light. Our G-MFO method strikes a training/validation/testing classification accuracy of $72.1\%/73.3\%/73.8\%$ when classifying $4$ types of WBC using a one-layer optical computing system, exceeding that of the HBT method (Fig.~\ref{fig: all-optical cell classfier}c). Furthermore, Fig.~\ref{fig: all-optical cell classfier}d shows that the inference enabled by optical computing is almost instantaneous ($\frac{d_{IO}+d_{OC}}{c}=1.4 \: ns$, where $c$ is the speed of light), compared to the $1.7\: ms$ of ResNet10, the electronic machine learning model used in \cite{wbcphase2021shu}. We need $1$ more milliseconds of in silico computing of region intensities corresponding to different classes to obtain the prediction. Such a step can be skipped if we use single-photon avalanche diode (SPAD)~\cite{bruschini2019single} point detectors to count the corresponding regions' cumulative signals. Though for now, the performance of our single-layer linear optical computing system is not on par with the electronic neural network, which hits a testing classification accuracy of $90.5\%$ \cite{wbcphase2021shu}, the potential of having ultra-high inference speed and $>70\%$ classification accuracy here point out an exciting direction on further increasing the complexity of our optical computing system to improve the absolute classification accuracy on classifying cells.

\section{Discussion}

\subsection{Future directions}

\subsubsection{Further apply G-MFO to more complex optical computing systems.}
Exploring the scalability of the optical computing system to more complex optical structures, along with integrating more layers and non-linear activation functions, could potentially enhance absolute performance. 

\subsubsection{Further improve the performance of G-MFO.}
Future research could also consider employing more advanced techniques related to Monte Carlo integration to reduce the training variance discussed in the previous Subsec.~\ref{subsec: G-MFO exhibits curse of dimensionality.}, which we anticipate could substantially broaden the viable search space, thus further empowering the G-MFO approach. These include using more advanced sampling strategies~\cite{caflisch1998monte} or integrating G-MFO with the SBT methods~\cite{kurenkov2021guiding} or adding critic function~\cite{konda1999actor}. The latter two methods trade off the introduced bias and sampling variance.

\subsubsection{Expanding the application of G-MFO to additional computational optics tasks.}
The concept of G-MFO presents promising avenues for application in other areas of computational optics, such as computer-generated holography~\cite{zhao2022model} and lens design~\cite{sitzmann2018end}. The inherent model-free and resource-efficient characteristics of G-MFO position it as a viable alternative to prevalent model-based methods~\cite{blinder2022state, sitzmann2018end}. Future research could focus on leveraging G-MFO to these domains, potentially enhancing computational efficiency and performance.

\subsection{Conclusion}
To conclude, our study underscores the effectiveness of a model-free strategy in training optical computing systems in situ, manifesting considerable potential in computational efficiency and reducing the simulation-to-reality performance gap. Although the study does not focus entirely on absolute image classification accuracy as it is based on a simple single or double diffractive optical computing system without non-linearity, it shows relative improvements compared to the existing training strategies, indicating that our strategy is a potentially valuable approach. The model-agnostic nature of our technique may become even more beneficial when implemented in intricate optical systems, representing a robust and versatile alternative to current strategies. It promises a strong foundation for exploring and practically implementing optical computing in real-world applications such as high-speed marker-free cell analysis.

%% file: sections/osa-supplemental-document-template.tex
\section*{Supplementary}
\renewcommand{\thefigure}{S.\arabic{figure}}
\setcounter{figure}{0}
\renewcommand{\thetable}{S.\arabic{table}}
\setcounter{section}{0}
\renewcommand{\thesection}{S.\arabic{section}}
\setcounter{subsection}{0}
\renewcommand{\thesubsection}{S.\arabic{subsection}}
\setcounter{table}{0}
\setcounter{equation}{0}
\setcounter{algorithm}{0}
\renewcommand{\theequation}{S.\arabic{equation}}
\renewcommand{\thealgorithm}{S.\arabic{algorithm}}

\vspace{0.3cm}
\vspace{0.5cm}
\section{Reproducibility of the hardware system} \label{subsec: Reproducibility}
\subsection{Hardware details of the optical computing system.}

\begin{figure}[h!]
\centering
\includegraphics[width=0.98\linewidth]{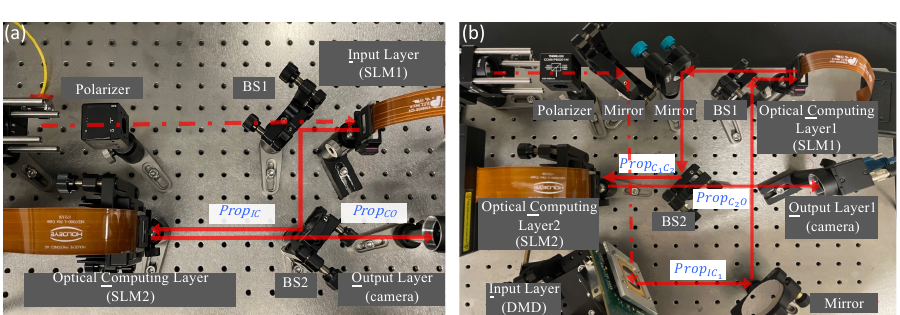}
\caption{\textbf{Real setups of our optical computing systems.} (a) Single-layer optical computing system. (b) Two-layer optical computing system.}
\label{fig:real_system}
\end{figure}

\begin{figure}[h!]
\centering
\includegraphics[width=0.98\linewidth]{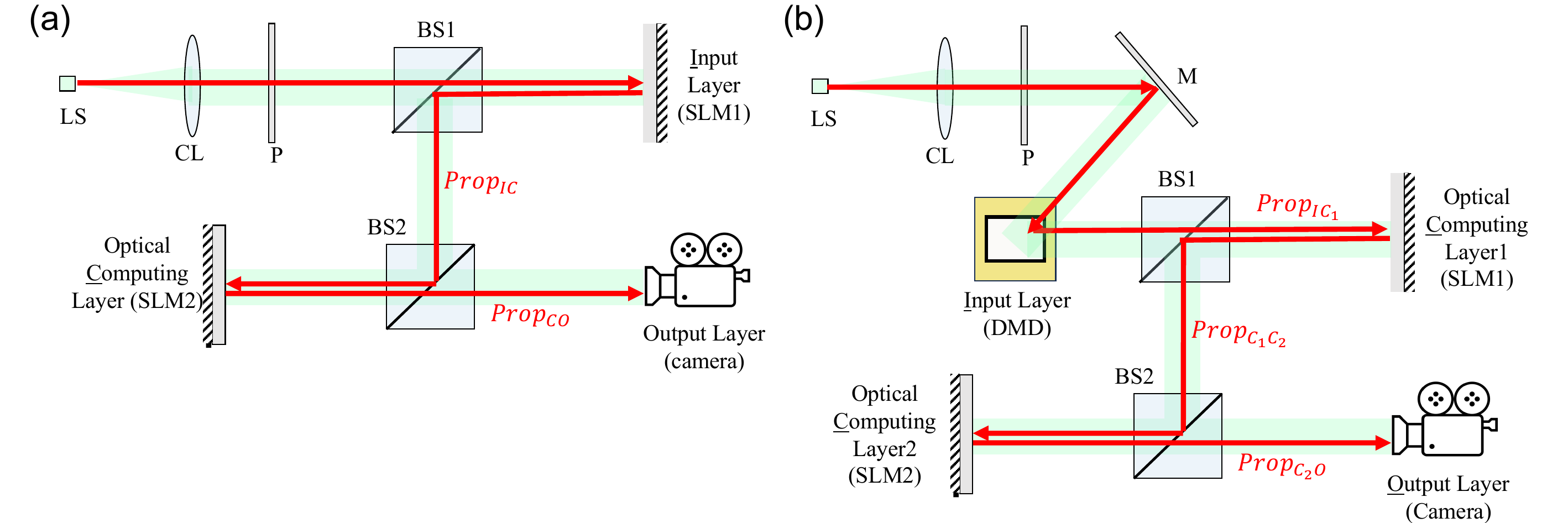}
\caption{\textbf{Setups of optical computing systems.} (a) In the single-layer optical computing system, we have a light source incident from the top left, passing through the collimator lens (CL), polarizer (P), and the first beam splitter (BS1) and then reflecting from the input layer, which displays phase object by phase delay from the SLM1. The reflected light then passes through BS1 and beam splitter 2 (BS2) and arrives at the optical computing layer, where we use SLM2 to upload the optical processing weights. The reflected light from the optical computing layer then arrives at the output layer, and we calculate the task-specific loss based on the arrival signal. We label the last two paths as $Prop_{IC}$ and $Prop_{CO}$, respectively, in the figure. (b) In the two-layer optical computing system, the light passing through CL and P is reflected by a mirror (M) and directed towards the input layer. The input layer displays binary intensity objects using a DMD. The reflected light then reaches the optical computing layer 1 and, subsequently, optical computing layer 2, where we use SLM1 and SLM2 to upload the optical processing weights. Finally, the light arrives at the output layer, and we obtain the output of the optical computing system. We label the three paths as $Prop_{IC_1}$, $Prop_{C_1C_2}$, $Prop_{C_2O}$.
}  
\verb||\label{fig: system skectch}
\end{figure}

\begin{figure}[h!]
\centering
\includegraphics[width=0.5\linewidth]{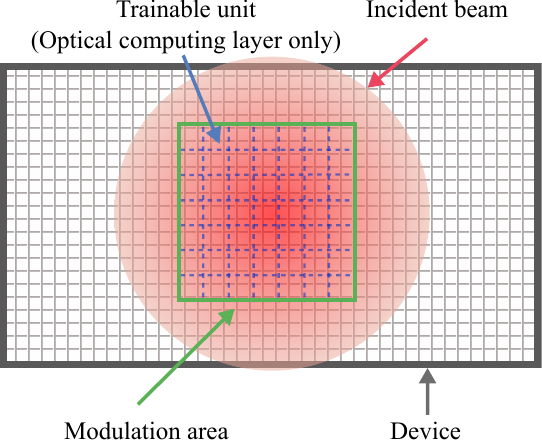}
\caption{\textbf{Optical computing modulation area.} In our experiments, the incident laser beam cannot fully illuminate the entire working area of the programmable optical devices and cameras. To ensure efficient modulation, we create a smaller optical computing modulation area for each device, fully covered by the incident beam. Pixel values outside this area are set to zero, allowing only those within the area to be configurable. 
Additionally, we divide the optical computing layer's modulation area into uniformly valued partitions. Each partition represents a trainable unit, which is also a trainable parameter, of $w$.
}
\label{fig:effective_shape}
\end{figure}

Fig.~\ref{fig:real_system} shows our home-built optical computing systems. Table.~\ref{tab:device_type} lists models of critical components used in our home-built optical computing systems. Table.~\ref{tab:device_param} further provides dimensions and frame rates of programmable optical devices (i.e., SLM1, SLM2 and DMD) and cameras involved in this work. The definition of the modulation area is shown in Fig.~\ref{fig:effective_shape}.

\subsection{Experimental setups of the real single-layer and two-layer optical computing systems $f_{sys-1-layer}$ and $f_{sys-2-layer}$. }~\label{} 

\textbf{Single-layer optical computing system:} A laser field $u_{laser}$ propagates onto the input layer (SLM1), where the phase-valued object $u_{obj}$ is displayed. The light field reflects from the input layer onto the optical computing layer $u_w$ (SLM2). A camera then detects the diffraction light at the output layer. The length $d_{IC}$ and $d_{CO}$ for $Prop_{IC}$ and $Prop_{CO}$ in Fig.~\ref{fig: system skectch}(a) are $215.1 mm$ and $201.6 mm$, respectively. As there are multiple components in the system (e.g., SLMs and cameras), it is difficult to manually align them with high accuracy. We developed a digital alignment method based on the homography technique~\cite{riba2020kornia} using SLMs to compensate for the alignment errors when we implement the baseline method HBT on the real single-layer optical computing system; otherwise, the performance will be pretty poor.
Details of our digital alignment method are in Supplement Sec.~\ref{subsec: calibration}.  

\textbf{Double-layer optical computing system:} A DMD displays the binary intensity object $u_{obj}$ on the input layer. The light field is then reflected from the input layer onto the optical computing layer1 $u_{w1}$ (SLM1) and subsequently to the optical computing layer2 $u_{w2}$ (SLM2). Following this, a camera captures the modulated light at the output layer. The length $d_{C_1C_2}$ and $d_{C_2O}$ for $Prop_{C_1C_2}$ and $Prop_{C_2O}$ in Fig.~\ref{fig: system skectch}(b) are $215.1 mm$ and $201.6 mm$, respectively. More details of the optical computing system are in Supplement Sec. S.1.

\subsection{Differentiable physics-based simulators $\hat{f}_{sys-1-layer}$ and $\hat{f}_{sys-2-layer}$.}\label{subsec: simulators} 

We construct an ideal physics-based simulator $\hat{f}_{sys}$ corresponding to the aforementioned optical computing system $f_{sys}$ as the sandbox to test different training algorithms. This simulator is also used inside the design loop of SBT and HBT, which serve as the baseline methods to compare.

Since our system only includes free-space wave propagation $\textcolor{red}{\hat{f}_{prop}}$, wavefront modulation $\textcolor{Colorfmod}{\hat{f}_{mod}}$, and sensor detection $\textcolor{Colorfdet}{\hat{f}_{det}}$, we build the optical computing simulator by stacking these three optical modules as building blocks.
The functions of optical modules are:
\begin{subequations} \label{eq: diff modules of system}
   \begin{align}
        &\textcolor{red}{\hat{f}_{prop}}(u_{in},z): u_{out} = \mathcal{F}^{-1} (\mathcal F(u_{in}) \times \mathcal F(h_{prop}(z))), \label{eq: free-space prop} \\
        &\textcolor{Colorfmod}{\hat{f}_{mod}}(u_{in}, u_{element}): u_{out} = u_{in} \times u_{element},\\
        &\textcolor{Colorfdet}{\hat{f}_{det}}(u_{in}):I_{out} = |u_{in}|^2, \label{eq: camera detection.}
    \end{align}
\vspace{0.2cm}
\end{subequations}
where $h_{prop}(z) = \frac {e^{jkz}}{j\lambda z}e^{\frac{jk}{2z}(x^2+y^2)}$ is the propagation kernel under Fresnel approximation~\cite{born2013principles} with a propagation distance $z$, wavelength $\lambda$ and angular wave number $k$, $u_{element}$ denotes the wavefront modulation from the programmable optical devices, $\mathcal{F}$ denotes the Fourier transform. 

\textit{Simulator of a single-layer optical computing system.} Based on the modules in Eq.~\ref{eq: diff modules of system}, the simulator of a single-layer $\hat{f}_{sys-1-layer} = \{\textcolor{Colorfmod}{\hat{f}_{mod}}, \textcolor{red}{\hat{f}_{prop}}, \textcolor{Colorfmod}{\hat{f}_{mod}}, \textcolor{red}{\hat{f}_{prop}}, \textcolor{Colorfdet}{\hat{f}_{det}}\}$ 
is constructed by chaining the building blocks ${1-5}$:
\begin{subequations}\label{eq: simuator}
    \begin{align}
        & 1. u_{out} = \textcolor{Colorfmod}{\hat{f}_{mod}}(u_{laser}, u_{obj}), \\
        & 2. u_{out} = \textcolor{red}{\hat{f}_{prop}}(u_{out}, d_{IC}), \\
        & 3. u_{out} = \textcolor{Colorfmod}{\hat{f}_{mod}}(u_{out}, u_{w}), \\
        & 4. u_{out} = \textcolor{red}{\hat{f}_{prop}}(u_{out}, d_{CO}), \\
        & 5. I_{cam} = \textcolor{Colorfdet}{\hat{f}_{det}}(u_{out}).
    \end{align} 
\end{subequations}

\textit{Simulator of a double-layer optical computing system} is constructed similarly by stacking layers as $\hat{f}_{sys-2-layer} = \{\textcolor{Colorfmod}{\hat{f}_{mod}}, \textcolor{red}{\hat{f}_{prop}}, \textcolor{Colorfmod}{\hat{f}_{mod}}, \textcolor{red}{\hat{f}_{prop}}, \textcolor{Colorfmod}{\hat{f}_{mod}}, \textcolor{red}{\hat{f}_{prop}}, \textcolor{Colorfdet}{\hat{f}_{det}}\}$:
\begin{subequations}\label{eq: simuator}
    \begin{align}
        & 1. u_{out} = \textcolor{Colorfmod}{\hat{f}_{mod}}(u_{laser}, u_{obj}), \\
        & 2. u_{out} = \textcolor{red}{\hat{f}_{prop}}(u_{out}, d_{IC_1}), \\
        & 3. u_{out} = \textcolor{Colorfmod}{\hat{f}_{mod}}(u_{out}, u_{w1}), \\
        & 4. u_{out} = \textcolor{red}{\hat{f}_{prop}}(u_{out}, d_{C_1C_2}), \\
        & 5. u_{out} = \textcolor{Colorfmod}{\hat{f}_{mod}}(u_{out}, u_{w2}), \\
        & 6. u_{out} = \textcolor{red}{\hat{f}_{prop}}(u_{out}, d_{C_2O}), \\   
        & 7. I_{cam} = \textcolor{Colorfdet}{\hat{f}_{det}}(u_{out}).
    \end{align} 
\end{subequations}

\vspace{0.3cm}
\subsection{Parameters of the optical computing system}
\begin{figure}[h!]
\centering
\includegraphics[width=0.65\linewidth]{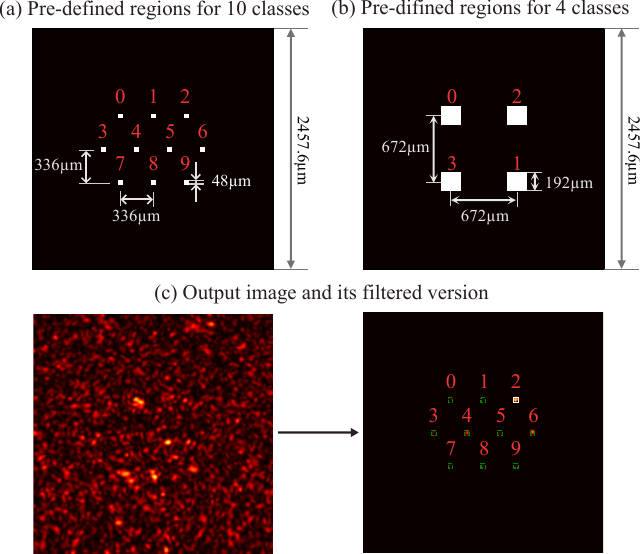}
\caption{\textbf{Getting classification prediction of the optical computing system.} We obtain the classification prediction of the optical computing system by taking an image with the camera and summing up the pixel values in different regions to get the intensity of each class. The class with the highest intensity is the predicted class of the system. (a) shows the pre-defined regions for 10 classes, we set the regions to have a size of $48 \mu m \times 48\mu m$ and the distance between two regions is $336 \mu m$. (b) shows pre-defined regions for 4 classes, we set the regions to have a size of $192 \mu m \times 192\mu m$ and the distance between two regions is $672 \mu m$. (c) shows an example of a captured image (lower left) and its filtered version (lower right). The filtered version sets pixel values outside the class regions to zero for clearer visualization.}
\label{fig:apply mask}
\end{figure}

We obtain the classification prediction of the optical computing system by comparing the intensities of different regions in the output layer~\cite{reconfigurable_onn, lin2018all}, as illustrated in Fig.~\ref{fig:apply mask}. 

\begin{table}[h!]
\centering
\begin{threeparttable}
\newcolumntype{P}[1]{>{\centering\arraybackslash}p{#1}}
\arrayrulecolor{black}
\renewcommand{\arraystretch}{1.0}
\begin{tabular}{P{0.22\linewidth} P{0.3\linewidth} P{0.28\linewidth}}
\hline
\textbf{Device} & \textbf{Part number} & \textbf{Manufacturer}\\
\hline
\rowcolor{gray!10}  SLM1        & PLUTO                & Holoeye\\
\rowcolor{white!10} SLM2        & GAEA                 & Holoeye\\
\rowcolor{gray!10}  DMD         & DLPLCR9000EVM       & TI\\
\rowcolor{white!10} Camera          & BFS-U3-13Y3C-C       & Teledyne FLIR\\
\rowcolor{gray!10}  BS              & HBS11-025-50-VIS     & Hengyang Optics\\
\rowcolor{white!10} Polarizer       & CCM5-PBS201/M        & Thorlabs\\
\hline
\end{tabular}
\caption{\textbf{Models of key components used in our optical computing system.}}
\label{tab:device_type}
\end{threeparttable}
\end{table}

\begin{table}[h!]
\centering
\begin{threeparttable}
\newcolumntype{P}[1]{>{\centering\arraybackslash}p{#1}}
\arrayrulecolor{black}
\renewcommand{\arraystretch}{1.0}
\begin{tabular} {P{0.12\linewidth} P{0.05\linewidth} P{0.19\linewidth} P{0.19\linewidth} P{0.15\linewidth}}
\hline
\textbf{Device} & \textbf{Pitch size ($\mu m$)} & \textbf{Full shape (pixels)} & \textbf{Modulation area shape (pixels)} & \textbf{Refreshing rate (FPS)} \\ \hline
\rowcolor{gray!10}  SLM1  & $8$       & $1080\times1920$     & $512\times512$      & $25$\tnote{*}\\
\rowcolor{white!10} SLM2   & $3.74$    & $2160\times3840$     & $1248\times1248$    & $30$\tnote{*}\\
\rowcolor{gray!10}  DMD        & $7.6$     & $1600\times2560$     & $512\times512$      & $125$\tnote{**}\\
\rowcolor{white!10} Camera     & $8$       & $1024\times1280$     & $512\times512$      & Max: $170$\\
\hline
\end{tabular}
\begin{tablenotes}
\item[*] We experimentally get 25 or 30 FPS in our self-written Python programs.
\item[**] The refresh rate of the DMD is limited by the camera's exposure time which is 4000 µs.
\end{tablenotes}
\caption{\textbf{Dimensions and frame rates of our experiments' programmable optical devices and cameras.}}
\label{tab:device_param}
\end{threeparttable}
\end{table}


\section{WBC dataset details}\label{sec: dataset}
The WBC phase image dataset \cite{wbcphase2021shu} has four classes of cells: granulocyte, monocyte, B cell, and T cell. Table~\ref{tab: WBC dataset} gives the number of cells in each class. 

\begin{table}[h!]
\centering
\newcolumntype{P}[1]{>{\centering\arraybackslash}p{#1}}
\arrayrulecolor{black}
\renewcommand{\arraystretch}{1.0}
\begin{tabular} {P{0.3\linewidth} P{0.15\linewidth} P{0.15\linewidth} P{0.15\linewidth}}
\hline
                & \textbf{Training  set} & \textbf{Validation set} &\textbf{Testing set}\\ \hline
\rowcolor{gray!10}  Granulocyte     & $465$      & $100$           & $100$\\
\rowcolor{white!10} Monocyte        & $513$      & $100$           & $100$\\
\rowcolor{gray!10}  B cell          & $437$      & $100$           & $100$\\
\rowcolor{white!10} T cell          & $330$      & $100$           & $100$\\
\rowcolor{gray!10}  Total           & $1745$     & $400$           & $400$\\
\hline
\end{tabular}
\caption{\textbf{Number of cells of each class in the WBC dataset.}}
\label{tab: WBC dataset}
\end{table}

\section{Training details} \label{subsec: training process}

\subsection{Computing resources.}
We evaluate our method in both simulation and experiments. The simulation runs on a high-end Linux server, while physical experiments are conducted on a Windows desktop. 
Table~\ref{tab:computer_configuration} shows the Windows desktop and Linux server configurations.

\begin{table}[h!]
\newcolumntype{P}[1]{>{\centering\arraybackslash}p{#1}}
\arrayrulecolor{black}
\centering
\renewcommand{\arraystretch}{1.0}
\begin{tabular}{P{0.3\linewidth} P{0.29\linewidth} P{0.27\linewidth}}
\hline
\toprule
                                     & \textbf{Server (Simulation)}                     & \textbf{Desktop (Experiments)}\\ \hline
\rowcolor{gray!10} CPU               & Intel Xeon Silver 4210R [$\times$2] & Intel Core i7-7700\\
\rowcolor{white!10}GPU               & Nvidia A6000 [$\times$2]            & Nvidia TITAN XP\\ 
\rowcolor{gray!10}GPU memory         & $48GB$ [$\times$2]                    & $12GB$\\ 
\rowcolor{white!10}RAM               & $192GB$                               & $32GB$\\ 
\rowcolor{gray!10}Operating system   & Ubuntu 18.04                        & Windows 10\\ 
\rowcolor{white!10}Graphic interface & No                                  & Yes\\
\bottomrule
\end{tabular}
\caption{\textbf{Configurations of our server and desktop.}}
\label{tab:computer_configuration}
\end{table}
\subsection{Training parameters and time of simulations and experiments.}

The training parameters and time of simulations and experiments are shown in Table.~\ref{tab: simulation and experiment parameters}. Unless otherwise noted, we early stop the training process when the validation loss does not decrease for 10 epochs in both simulations and real system experiments. We choose the checkpoint with the lowest validation loss and use it to evaluate the test set. 

In Sec. 4.1.1 of the main text, the number of training epochs for G-MFO is set to $500$ epochs, and the number of training epochs for the two-point optimization method is set to $6400$ epochs. In contrast, the number of training epochs for the full-point optimization method is limited to $100$ due to its much longer training time per epoch. Despite these differences, all three experiments converge well by the end of training. We take the training/validation/testing accuracy in the checkpoint with the lowest training loss as the result.

Due to the extensive time required for G-MFO simulations and experiments involving 10,000 training data on the two-layer optical computing system in Sec. 4.1.4, we stop these training processes at the $10th$ epoch and $9th$ epoch, respectively.

In Sec. 4.3 of the main text, we assess the G-MFO method's potential for overfitting using a single-layer optical computing system simulator. Thus, we stop the training process at the 200th epoch or after 50 epochs of no decrease in the training loss. We take the training accuracy in the checkpoint with the lowest training loss as the result for each simulation in Sec. 4.3.

\begin{table}[h!]
\newcolumntype{P}[1]{>{\centering\arraybackslash}p{#1}}
\begin{threeparttable}
\arrayrulecolor{black}
\centering
\renewcommand{\arraystretch}{1.0}
\begin{tabular} {P{0.1\linewidth} P{0.08\linewidth} P{0.05\linewidth} P{0.05\linewidth} P{0.05\linewidth} P{0.05\linewidth} P{0.1\linewidth} P{0.1\linewidth} P{0.08\linewidth} P{0.1\linewidth}}
\hline
    & Number of trainable parameters & Sampling size & Batch size  & Optimizer & Learning rate & \textit{In silico} computation time\tnote{*} & Device waiting time for getting $f_{sys}(x,w)$\tnote{*} & Total training time\tnote{*} & Stop criteria \\ \hline
\rowcolor{gray!10} \multicolumn{10}{c}{MNIST and FMNIST simulations on the two-layer optical computing system $\hat{f}_{sys-2-layer}$}\\
\rowcolor{gray!10} \multicolumn{10}{c}{(\textit{$10,000$ data points each for training, validation and testing).}}\\
Ideal & $128^2$ & None  & $32$ & SGD & $10$ & $1m47s$ & None & $1m47s$ & ES-Val-10\\
HBT   & $128^2$ & None  & $32$ & SGD & $10$ & $2m25s$  & None & $2m25s$ & ES-Val-10\\
G-MFO   & $128^2$ & $128$ & $32$ & SGD & $20$ & $6h2m25s$ & None & $6h2m25s$ & $10th$ epoch\\
\rowcolor{gray!10} \multicolumn{10}{c}{MNIST and FMNIST experiments on the two-layer optical computing system $f_{sys-2-layer}$}\\
\rowcolor{gray!10} \multicolumn{10}{c}{(\textit{$10,000$ data points each for training, validation and testing).}}\\
G-MFO   & $128^2$ & $128$  & $32$ & SGD & $20$ & $24m53s$ & $5h38m20s$ & $6h3m13s$ & $9th$ epoch\\
\rowcolor{gray!10} \multicolumn{10}{c}{MNIST and FMNIST simulations on the single-layer optical computing system $\hat{f}_{sys-1-layer}$}\\
\rowcolor{gray!10} \multicolumn{10}{c}{(\textit{$1,000$ data points each for training, validation and testing).}}\\
Ideal \& SBT & $128^2$ & None & $32$ & Adam & $0.01$ & $17s$ & None & $17s$ & ES-Val-10\\
G-MFO          & $128^2$ & $128$& $32$ & SGD  & $15$ & $11m58s$ & None & $11m58s$ & ES-Train-50 or $200th$ epoch\\
\rowcolor{gray!10} \multicolumn{10}{c}{FMNIST simulations on the single-layer optical computing system $\hat{f}_{sys-1-layer}$}\\
\rowcolor{gray!10} \multicolumn{10}{c}{(\textit{$100$ data points each for training, validation and testing).}}\\
Two-point & $128^2$ & 2 & $32$ & SGD & $2$ & $9s$ & None & $9s$ & $6400th$ epoch\\
Full-point & $128^2$ & $128^2$ & $32$ & SGD & $20$ & $1h53m$ & None & $1h53m$ & $100th$ epoch\\
G-MFO      & $128^2$ & $128$& $32$ & SGD  & $20$ & $2m$ & None & $2m$ & $500th$ epoch\\
\rowcolor{gray!10} \multicolumn{10}{c}{MNIST and FMNIST experiments on the single-layer optical computing system $f_{sys-1-layer}$}\\
\rowcolor{gray!10} \multicolumn{10}{c}{(\textit{$1,000$ data points each for training, validation and testing).}}\\
HBT   & $128^2$ & None  & $4$ & Adam & $0.01$ & $2m16s$ & $2m18s$ & $4m34s$ & ES-Val-10\\
G-MFO   & $128^2$ & $128$ & $32$ & SGD & $100$ & $6s$ & $1h56m1s$ & $1h56m7s$ & ES-Val-10\\
HBT+G-MFO   & $128^2$ & $128$ & $32$ & SGD & $10$ & $6s$ & $1h56m1s$ & $1h56m7s$ & ES-Val-10\\
\rowcolor{gray!10} \multicolumn{10}{c}{WBC experiments on the single-layer optical computing system $f_{sys-1-layer}$}\\
HBT   & $128^2$ & None  & $4$ & Adam & $0.005$ & $4m4s$ & $2m6s$ & $6m15s$ & ES-Val-10\\
G-MFO   & $128^2$ & $64$ & $32$ & SGD & $50$ & $1m11s$ & $1h42m17s$ & $1h43m28s$ & ES-Val-10\\
\hline
\end{tabular}
\begin{tablenotes}
\item[*] Time per epoch.
\end{tablenotes}
\caption{\textbf{Training parameters and time for experiments and simulations.} ES-Val-10 denotes stopping training after 10 epochs without a decrease in validation loss, while ES-Train-50 denotes stopping training following 50 epochs of no decrease in training loss.}
\label{tab: simulation and experiment parameters}
\end{threeparttable}
\end{table}

\subsection{Training process for experiments.}

The training process for G-MFO and HBT experiments consists of getting optical computing system output $f_{sys}(x,w)$. Experiment workflow.~\ref{workflow: HBT} shows how to get $\{f_{sys}(x_i, w)\}_{i=1}^B$ in the HBT experiment. 
In the G-MFO experiment, we use different experiment workflows to get $\{f_{sys}(x_i, w_j)\}_{i=1, j=1}^{B, M}$ depending on the specific devices used as input and optical computing layers. In experiments on the single-layer optical computing system ($responsetime_I>responsetime_O$), we use Experiment workflow~\ref{workflow: MFO mask} for the sake of a shorter waiting time. In experiments on the two-layer optical computing system  ($responsetime_I<responsetime_O$), we use Experiment workflow~\ref{workflow: MFO input}. $B$ is the batch size, $M$ is the sampling size, $responsetime_I$ is the input layer device response time and $responsetime_O$ is the optical computing layer device response time.

\begin{algorithm}[h!]
\floatname{algorithm}{Experiment workflow}
\caption{Acquiring $\{f_{sys}(x_i, w)\}_{i=1}^B$ in HBT experiment.}\label{workflow: HBT}
\begin{algorithmic}[1]
\State \textbf{Input: }  {A batch of classification dataset $\mathcal{D}_b=\{x_i, y_i\}_{i=1}^B$ with batch size $B$, optical computing weight $w$, optical computing system $f_{sys}$, input layer device response time $responsetime_I$, optical computing layer device response time $responsetime_O$.}
\State \textbf{Output: }{Optical computing system output $\{f_{sys}(x_i, w)\}_{i=1}^B$. 
}
\State{Refresh $w$ on the optical computing layer.}
\State{Wait $responsetime_O$.}  
\For{$i$ in range $B$}
    \State{Display $x_i$ on input layer.}
    \State{Wait $responsetime_I$.}
    \State{Capture $f_{sys}(x_i, w)$ with camera.}
\EndFor
\State{\textbf{Total waiting time}: $responsetime_O + B \times responsetime_I$.}
\end{algorithmic}
\vspace{-0.1cm}
\end{algorithm}

\begin{algorithm}[h!]
\floatname{algorithm}{Experiment workflow}
\caption{Acquiring $\{f_{sys}(x_i, w_j)\}_{i=1, j=1}^{B, M}$ in G-MFO experiment when optical computing layer device has a short response time.}\label{workflow: MFO mask}
\begin{algorithmic}[1]
\State \textbf{Input:} {A batch of classification dataset $\mathcal{D}_b=\{x_i, y_i\}_{i=1}^B$ with batch size $B$, a group of sampled optical computing weights $\{w_j\}_{j=1}^M$, optical computing system $f_{sys}$, input layer device response time $responsetime_I$, optical computing layer device response time $responsetime_O$.}
\State \textbf{Output:} {Optical computing system output $\{f_{sys}(x_i, w_j)\}_{i=1, j=1}^{B, M}$.}
\For{$i$ in range $B$}
    \State{Display $x_i$ on input layer.}
    \State{Wait $responsetime_I$.}
    \For{$j$ in range $M$}
        \State{Refresh $w_j$ on the optical computing layer.}
        \State{Wait $responsetime_O$.}        
        \State{Capture $f_{sys}(x_i, w_j)$ with camera.}
    \EndFor
\EndFor
\State{\textbf{Total waiting time}: $B \times M \times responsetime_O + B \times responsetime_I$}
\end{algorithmic}
\vspace{-0.1cm}
\end{algorithm}

\begin{algorithm}[!h]
\floatname{algorithm}{Experiment workflow}
\caption{Acquiring $\{f_{sys}(x_i, w_j)\}_{i=1, j=1}^{B, M}$ in G-MFO experiment when input layer device has a short response time.} \label{workflow: MFO input}
\begin{algorithmic}[1]
\State \textbf{Input: }  {A batch of classification dataset $\mathcal{D}_b=\{x_i, y_i\}_{i=1}^B$ with batch size $B$, a group of sampled optical computing weights $\{w_j\}_{j=1}^M$, optical computing system $f_{sys}$, input layer device response time $responsetime_I$, optical computing layer device response time $responsetime_O$.}
\State \textbf{Output: }{Optical computing system output $\{f_{sys}(x_i, w_j)\}_{i=1, j=1}^{B, M}$.}
\For{$j$ in range $M$}
    \State{Refresh $w_j$ on the optical computing layer}
    \State{Wait $responsetime_O$}
    \For{$i$ in range $B$}
        \State{Display $x_i$ on input layer}
        \State{Wait $responsetime_I$}        
        \State{Capture $f_{sys}(x_i, w_j)$ with camera}
    \EndFor
\EndFor
\State{\textbf{Total waiting time}: $M \times responsetime_O + B \times M \times responsetime_I$}
\end{algorithmic}
\vspace{-0.1cm}
\end{algorithm}

\section{Compare the overall training time of G-MFO and HBT}~\label{subsec: time bottleneck from refreshing time}
The maximum frame rate for the phase-SLM is approximately 30 FPS, while the DMD can reach approximately 125 FPS which is limited by the camera's exposure time (refer to Table~\ref{tab:device_param}). These low refresh rates more adversely affects the training duration of the G-MFO compared to the HBT, with the impact scaling in relation to the G-MFO's sampling size (refer to the device waiting time for computing $f_{sys}(x,w)$ in Table~\ref{tab: simulation and experiment parameters}).
This time bottleneck can be alleviated when substituting the conventional phase-SLM with the fast ones, such as the Heavily-quantized Spatial Light Modulators~\cite{choi2022time} or enhance the light intensity to reduce the camera's exposure time, thereby increasing the DMD frame rate.

\section{Calibration of the system} \label{subsec: calibration}
The calibration of the optical computing system is critical for reproducing baseline methods that rely on the simulator $\hat{f}_{sys}$ as tiny misalignment between multiple planes will hugely degrade the performance, as we have shown in Fig. 2 in the main text. 
Our digital alignment for a single-layer optical computing system consists of estimating $2$ sets of parameters: lengths $\{d_{IO}, d_{CO}\}$ and homography transformations $\{H_{IO}, H_{CO}\}$. 
Specifically, $H_{IO}$ is the projective transformation between the input layer and output layer, $d_{IO}$ is the corresponding distance, $H_{CO}$ is the projective transformation between the optical computing layer and output layer, and $d_{CO}$ is the corresponding distance.

\subsection{Calibrate propagation distance z using holography.}
We first use free-space holography~\cite{blinder2022state} to calibrate the distance $d$ between planes. We concatenate the building blocks Eq.~\ref{eq: free-space prop} and ~\ref{eq: camera detection.} and construct a free-space holography simulator $\hat{f}_{holo}=\{\hat{f}_{prop}, \hat{f}_{det}\}$. 
The input to the holography simulator is $u_{in}=e^{j\phi_{holo}}$ and $z=d$, where $\phi_{holo}$ is the pre-calculated phase that generates the hologram, while $d$ is the distance to optimize. 
Calculation of $\phi_{holo}$ is achieved by extracting the phase of Fourier transform of the target obj $I_{target}$:
\begin{equation}
    \phi_{holo} = \text{angle}(\mathcal{F}(\sqrt{I_{target}}))
\end{equation}
Then we upload the $\phi_{holo}$ onto the real SLM with the Holoeye SLM Pattern Generator software, add various defocus phase masks corresponding to various $d$, visually compare the quality of the output hologram and select the best fit $d$ as the calibrated distance.  

\subsection{Homography estimation among multiple planes.}
Homography is a linear transformation between corresponding points in
two images with an $8$ degree of freedom~\cite{riba2020kornia}. It is solved by minimizing dense photometric loss or by feature matching. 
Our experiment has two SLMs and a camera (Fig.~\ref{fig: system skectch}(a)). Since directly calibrating $H_{IC}$, the transformation between the two SLMs, is infeasible as we cannot put the camera on any plane of these two SLMs; we calibrate their transformations to the output plane (the camera plane), respectively. Take the estimation  of photometric-based homography between the input layer and the output layer as an example. The objective function is:
\begin{equation}
    H_{IO}^{\star} =  \operatorname*{arg\,min}_{H_{IO}} \mathcal{L}_{homo}(\sqrt{I_{target}}-warp(|\hat{f}_{prop}(u_{in}, d_{IO})|),
\end{equation}
where $warp$ is the homography warpper given the homography estimation $H_{IO}$. We solve this objective function with auto-differentiation~\cite{riba2020kornia}.

\subsection{Rebuild the simulator with homography estimation in the loop.} 
In the previous subsection, we discuss estimating the homography transformation between planes in the real system.

\begin{figure}[h!]
\centering
\includegraphics[width=0.5\linewidth]{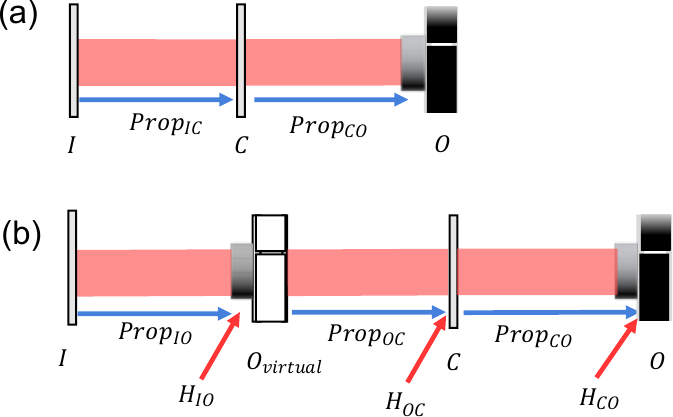}
\caption{We rebuild the original simulator as sketched in (a) into the three-propagator-based simulator in (b) to incorporate the homography estimations into the simulator. $O_{virtual}$ denotes the virtual camera plane, which does not exist in the real system. Here we use $O_{virtual}$ to help calibrate the system.}
\label{fig: three_virtual_prop}
\end{figure}

To incorporate the calibrated homography matrices into our system simulator $\hat{f}_{sys}$, we rebuild the simulator with three virtual propagators instead of two. The latter is discussed in Subsec. S1.3 and depicted in Fig~\ref{fig: three_virtual_prop}(a). We use a 3-layer simulator because the homography wrapper $warp$ and the wave propagator $\hat{f}_{prop}$ are not commutative. In other words:
\begin{equation}
    warp(\hat{f}_{prop}(u_{in}, d_{IO}), H_{IO}) \neq \hat{f}_{prop}(warp(u_{in}, H_{IO}), d_{IO}). 
\end{equation}

A sketch of the rebuilt simulator is in Fig.~\ref{fig: three_virtual_prop}(b). We introduce a virtual output plane $O_{virtual}$ into the simulator so that we can incorporate the homography estimations $H_{IO}$ and $H_{CO}$ into the simulation loop. Since the homography matrix is estimated at the output planes of the optical propagator, the settings in Fig.~\ref{fig: three_virtual_prop}(b) enable the use of homography matrix at the output planes of the propagator, which cannot be done via using the original simulator in Fig.~\ref{fig: three_virtual_prop}(a). 

\section{Baseline method -- Hybrid training.} \label{subsec: baseline}

We include the hybrid training (HBT)/physics-aware training as one of the baseline methods to compare~\cite{spall2022hybrid, wright2022deep}.
The philosophy of hybrid training is simple: when updating the optical computing system, we use the real system to do the forward pass to get the value of the objective function $J(w)$:
\begin{equation}
    J(w) = \frac{1}{N}\sum_{i=1}^{N}\mathcal{L}(f_{sys}(x_i; w), y_i).
\end{equation}

In the backward update pass, we substitute the $f_{sys}$ with its differentiable simulator $\hat{f}_{sys}$ and update the weights $w$ via:
\begin{equation}
    w =  w + \alpha \frac{\partial J(w)}{\partial \hat{f}_{sys}} \frac{\partial \hat{f}_{sys}}{w}.
\end{equation}

This trick enables the backward pass in the biased while differentiable simulator $\hat{f}_{sys}$. The critical difference between the hybrid training (HBT) method and the simulator-based training (SBT) is that the former does the forward pass in the real system ${f}_{sys}$ while the latter conducts both passes in the simulator ${\hat{f}}_{sys}$.

\begin{figure*}[t]
\newcolumntype{P}[1]{>{\centering\arraybackslash}p{#1}}
\centering
\includegraphics[width=0.98\linewidth]{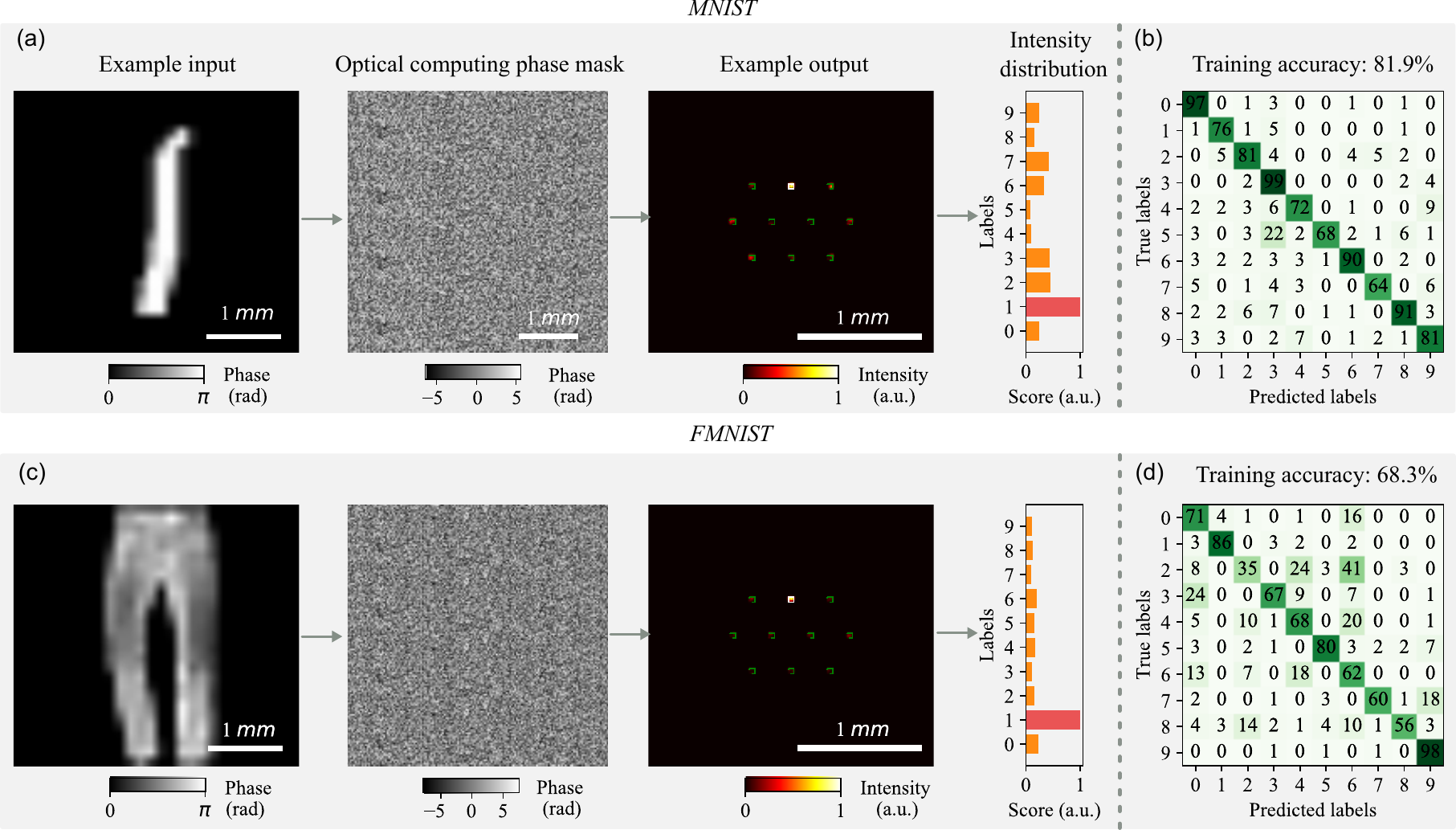}
\caption{\textbf{Example outputs and confusion matrices of the single-layer optical computing system trained with HBT optimization.}
(a) the optical computing system trained with HBT correctly predicts an input phase object digit ‘$1$’ from the MNIST dataset. In (b), we show we achieve an accuracy of $81.9\%$ on the MNIST training dataset. (c) an example of ‘trouser’ from the FMNIST training dataset is correctly predicted. (d) shows that the accuracy on the FMNIST training dataset is $68.3\%$.}
\verb||\label{fig: HBT result}
\end{figure*}

We visualize some experimental outputs and confusion matrices using HBT on a single-layer optical computing system in Fig.~\ref{fig: HBT result}.

\rev{\section{Comparision between G-MFO and L-MBO.}}

\rev{In related work, Sec. 2.1, we discuss the \textit{hardware-in-the-loop} or \textit{real2sim}-style algorithm L-MBO. In this section, we compare L-MBO and our method in terms of performance.}

\rev{\subsection{Compare G-MFO with our implementation of L-MBO.}}

\rev{We implement an L-MBO method and compare it with G-MFO on a single-layer optical computing simulator $\hat{f}_{sys-1-layer}$ using the MNIST dataset. We use $1,000$ data to train, $1,000$ data to validate, and $1,000$ data to test.
Our implementation of L–MBO on the simulator consists of:} 

\begin{itemize}
    \item Feeding the task dataset into the simulator to get an exploration dataset for training the system;
    \item Pre-training a physics-based learned proxy of a misaligned system through supervised learning;
    \item Optimizing the task performance based on the pre-trained proxy;
    \item Testing the optimized optical weights in the simulator to get train/validation/testing accuracy.
\end{itemize}

\rev{Specifically, as for the physics-based learned proxy, the model is constructed by chaining a UNet ~\cite{ronneberger2015u} with a physics-based simulator of a misalignment system. We do not use pure network-based proxy as the overfitting is evident without physics prior.}

\rev{We present the simulation results in Table~S.6.  In our L-MBO implementation, the L-MBO under-performs our method; G-MFO’s train accuracy is $90.4\%$ while the L-MBO’s is $80.2\%$. We attribute this to multiple reasons, such as not including alternating training during the task optimization process~\cite{zheng2023dual}, the proxy’s hyper-parameters not being well-tuned, etc.}

\begin{table}[h!]
\newcolumntype {C}[1]{>{\centering\arraybackslash}m{#1}}
\arrayrulecolor{black}
\centering
\renewcommand{\arraystretch}{1.5}
\begin{tabular}{C{0.16\linewidth} | C{0.08\linewidth} | C{0.08\linewidth}| C{0.08\linewidth}}
\hline
             &\multicolumn{3}{c}{Accuracy}    \\\hline
             & Train             & Val                 & Test \\\hline\hline
L-MBO        & $80.2\%$          & $69.8\%$            & $65.2\%$ \\\hline
G-MFO (Ours) & $90.4\%$          & $83.5\%$            & $81.5\%$\\\hline
\end{tabular}
\caption{\rev{\textbf{Numerical comparison of G-MFO and L-MBO (our implementation) on the MNIST dataset using a simulated single-layer diffractive optical computing system. 
} }}
\label{tab: l-mbo comparison result}
\end{table}

\rev{\subsection{Compare G-MFO with a reported L-MBO result.}}

\rev{Note that in an experimental realization of a 2-layer linear diffractive optical computing system, L-MBO's\cite{zheng2023dual} reported testing accuracy ($87.5\%$) is better than ours ($80.3\%$), but their code for the diffractive optical computing system has not been open-sourced yet.} 

Moreover, as highlighted in Table 1, our method has advantages: 
\begin{itemize}
    \item Ours is memory- and computation-efficient.
    \item  Ours does not require collecting high-fidelity images of input objects, which are costly to acquire in terms of instruments, time, and electronic storage.
\end{itemize}

\rev{We anticipate further investigation and integration of these two methods would be an interesting research.}